\documentclass[sigconf]{acmart}

\AtBeginDocument{%
  }

\usepackage{booktabs}
\usepackage{longtable}
\usepackage{xcolor}
\usepackage{bm}
\usepackage{multirow}
\usepackage{graphicx}
\usepackage{makecell}
\usepackage{subcaption}
\usepackage{lipsum}
\usepackage{enumitem}
\usepackage{amsmath}
\usepackage{amsfonts}
\usepackage{threeparttable}
\usepackage{soul}

\newcommand{\upwardbf}[1]{\textcolor{red}{\bm{$\uparrow$}\textbf{#1}}}
\newcommand{\upcell}[2]{\makecell{#1\\($\uparrow${#2\%})}}
\newcommand{\downcell}[2]{\makecell{#1\\($\downarrow${#2\%})}}

\acmISBN{978-1-4503-XXXX-X/2018/06}

\begin{document}

\title{Think Before Recommend: Unleashing the Latent Reasoning Power for Sequential Recommendation}


\author{Jiakai Tang\textsuperscript{1}\textsuperscript{$\ast$}\textsuperscript{$\mathsection$}, Sunhao Dai\textsuperscript{1}\textsuperscript{$\ast$}, Teng Shi\textsuperscript{1}, Jun Xu\textsuperscript{1}, Xu Chen\textsuperscript{1}\textsuperscript{$\dagger$}, Wen Chen\textsuperscript{2}\textsuperscript{$\dagger$}, Jian Wu\textsuperscript{2}, Yuning Jiang\textsuperscript{2}}

\affiliation{%
  \institution{\textsuperscript{1}Gaoling School of Artificial Intelligence, Renmin University of China, Beijing, China}
  \city{\textsuperscript{2}Alibaba Group, Beijing, China}
  \country{}
}
\email{{tangjiakai5704,sunhaodai,shiteng,junxu,xu.chen}@ruc.edu.cn}
\email{{chenyu.cw,joshuawu.wujian,mengzhu.jyn}@alibaba-inc.com}
\thanks{$\ast$ Equal Contribution.}
\thanks{$\mathsection$ Work done during internship at Alibaba Group.}
\thanks{$\dagger$ Corresponding author.}

\renewcommand{\authors}{Jiakai Tang, Sunhao Dai, Teng Shi, Jun Xu, Xu Chen, Wen Chen, Jian Wu, Yuning Jiang}
\renewcommand{\shortauthors}{Jiakai Tang et al.}
\renewcommand{\shorttitle}{Think Before Recommend: Unleashing the Latent Reasoning Power for Sequential Recommendation}

\begin{abstract}
Sequential Recommendation (SeqRec) aims to predict the next item by capturing sequential patterns from users' historical interactions, playing a crucial role in many real-world recommender systems.
However, existing approaches predominantly adopt a direct forward computation paradigm, where the final hidden state of the sequence encoder serves as the user representation. We argue that this inference paradigm, due to its limited computational depth, struggles to model the complex evolving nature of user preferences and lacks a nuanced understanding of long-tail items, leading to suboptimal performance.
To address this issue, we propose \textbf{ReaRec}, the first inference-time computing framework for recommender systems, 
which enhances user representations through implicit multi-step reasoning.
Specifically, ReaRec autoregressively feeds the sequence's last hidden state into the sequential recommender while incorporating special reasoning position embeddings to decouple the original item encoding space from the multi-step reasoning space.
Moreover, we introduce two lightweight reasoning-based learning methods, Ensemble Reasoning Learning (ERL) and Progressive Reasoning Learning (PRL), to further effectively exploit ReaRec's reasoning potential.
Extensive experiments on five public real-world datasets and different SeqRec architectures demonstrate the generality and effectiveness of our proposed ReaRec.
Remarkably, post-hoc analyses reveal that ReaRec significantly elevates the performance ceiling of multiple sequential recommendation backbones by approximately 30\%-50\%.
Thus, we believe this work can open a new and promising avenue for future research in inference-time computing for sequential recommendation.

\end{abstract}




\maketitle

\section{Introduction}

\begin{figure}
  \centering
  \includegraphics[width=0.85\linewidth]{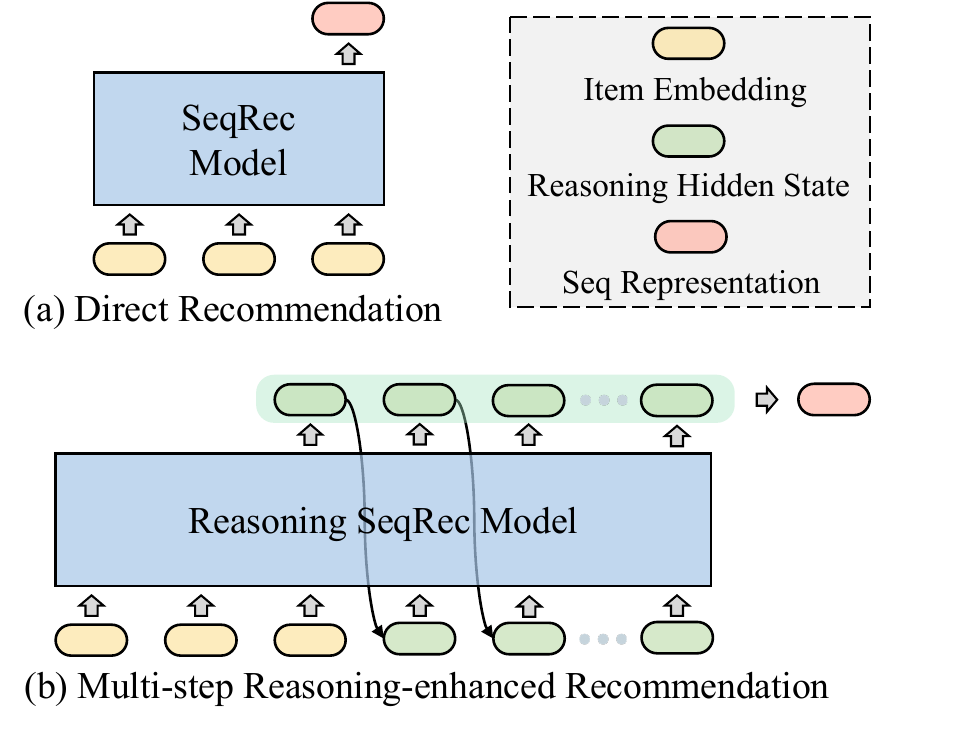}
  \caption{Illustration of traditional direct inference (\textit{i.e.}, reasoning-free) and our proposed multi-step reasoning-enhanced sequential recommendation framework.}
  \label{fig:intro}
\end{figure}

Recommender systems (RS) have become ubiquitous in modern daily life, powering personalized services across domains such as e-commerce platforms~\cite{zhou2018deep,singer2022sequential}, music recommendation services~\cite{dai2024modeling,zhang2022counteracting}, and video streaming applications~\cite{lei2021semi,zhang2024saqrec}. To accurately capture a user's next interaction intent, sequential recommendation algorithms are designed to analyze historical interactions to mine underlying sequential patterns and model latent user preferences~\cite{boka2024survey,wang2019sequential,fang2020deep}. Current mainstream sequential recommendation models, such as SASRec~\cite{kang2018self} and UniSRec~\cite{hou2022towards}, adopt a Transformer-based architecture, leveraging their power attention mechanisms to adaptively weight past interacted items and use the final position's encoded output as the user representation, as illustrated in Fig.~\ref{fig:intro}(a). However, we argue this prevailing direct forward inference paradigm may lack nuanced comprehension of dynamic user preferences and evolving interest patterns, leading to suboptimal modeling for long-tail user interest and unpopular items.
Despite their efficiency, we argue that these direct inference paradigms often fall short in modeling long-tail users with fewer interactions and less popular items—scenarios that inherently demand more nuanced reasoning and deeper representation learning.

Recently, many studies from the natural language processing (NLP) community have demonstrated that \textit{Chain-of-Thought (CoT)} during inference can significantly improve the performance of \textit{Large Language Models (LLMs)} on complex tasks like mathematics and coding~\cite{shao2024deepseekmath,guo2024deepseek,team2025kimi, wei2022chain}.
By allowing the model to perform multi-step deliberation before generating a final output, CoT-based reasoning enhances the model's capacity to handle complex problems beyond what direct inference allows.
Furthermore, \citet{feng2023towards}~theoretically uncover that the emergent thinking capabilities are attributed to the increased computational depth introduced by CoT-based reasoning, which allows models to overcome the expressivity limitations of direct answer even with constrained parameter sizes.

Motivated by these insights, we explore whether a similar \textit{think-before-action} paradigm can benefit sequential recommendation, especially for challenging cases such as long-tail users and items. We propose \textbf{\textit{ReaRec}}, a novel reasoning-enhanced framework that enables SeqRec models to engage in implicit multi-step reasoning during inference. 
As shown in Fig.~\ref{fig:intro}(b), ReaRec performs autoregressive reasoning over latent representations before producing the final user embedding, thereby deepening feature crossing and improving representational richness. 
To prevent the recommender from confusing the sequence encoding stage and reasoning stage, we design a specialized positional encoding scheme to explicitly distinguish item representations from reasoning inputs. 
However, unlike NLP tasks, where explicit reasoning chains naturally provide process supervision to guide model optimization~\cite{lightman2023let,luo2024improve,setlur2024rewarding}, implicit reasoning in sequential recommendation lacks effective intermediate signals. This absence of stepwise guidance could lead to unpredicted \textit{reasoning degradation} issues, causing the recommender to either replicate prior reasoning patterns or progressively drift away from accurately modeling the user's true interest distribution. Consequently, this may significantly impair the robustness and generalization capability of the recommendation model.

To address the aforementioned challenges, we propose two simple yet effective reasoning learning strategies, \textbf{\textit{Ensemble Reasoning Learning (ERL)}} and \textbf{\textit{Progressive Reasoning Learning (PRL)}}, to fully exploit the reasoning power of our ReaRec framework.
\underline{For the ERL method}, it leverages the idea of \textit{ensemble learning} to construct multi-order user representations to comprehensively capture latent interest distributions from diverse perspectives. Specifically, we introduce multi-step supervised optimization to alleviate the optimization difficulty in deep reasoning processes. Furthermore, to prevent reasoning-pattern degradation, we incorporate a representation diversity regularizer to mitigate output homogeneity in multi-step reasoning.
\underline{For the PRL method}, inspired by \textit{curriculum learning}, we design a progressive temperature annealing mechanism to guide the model from initial exploitation to the gradual refinement of modeled sequential patterns. This approach enables the model to progressively learn the user's true interest distributions. Moreover, we also propose a reasoning-aware contrastive learning objective to enhance the reasoning robustness ability by simulating the error self-correction process, thus achieving better generalization performance.

\begin{figure}
  \centering
  \includegraphics[width=\linewidth]{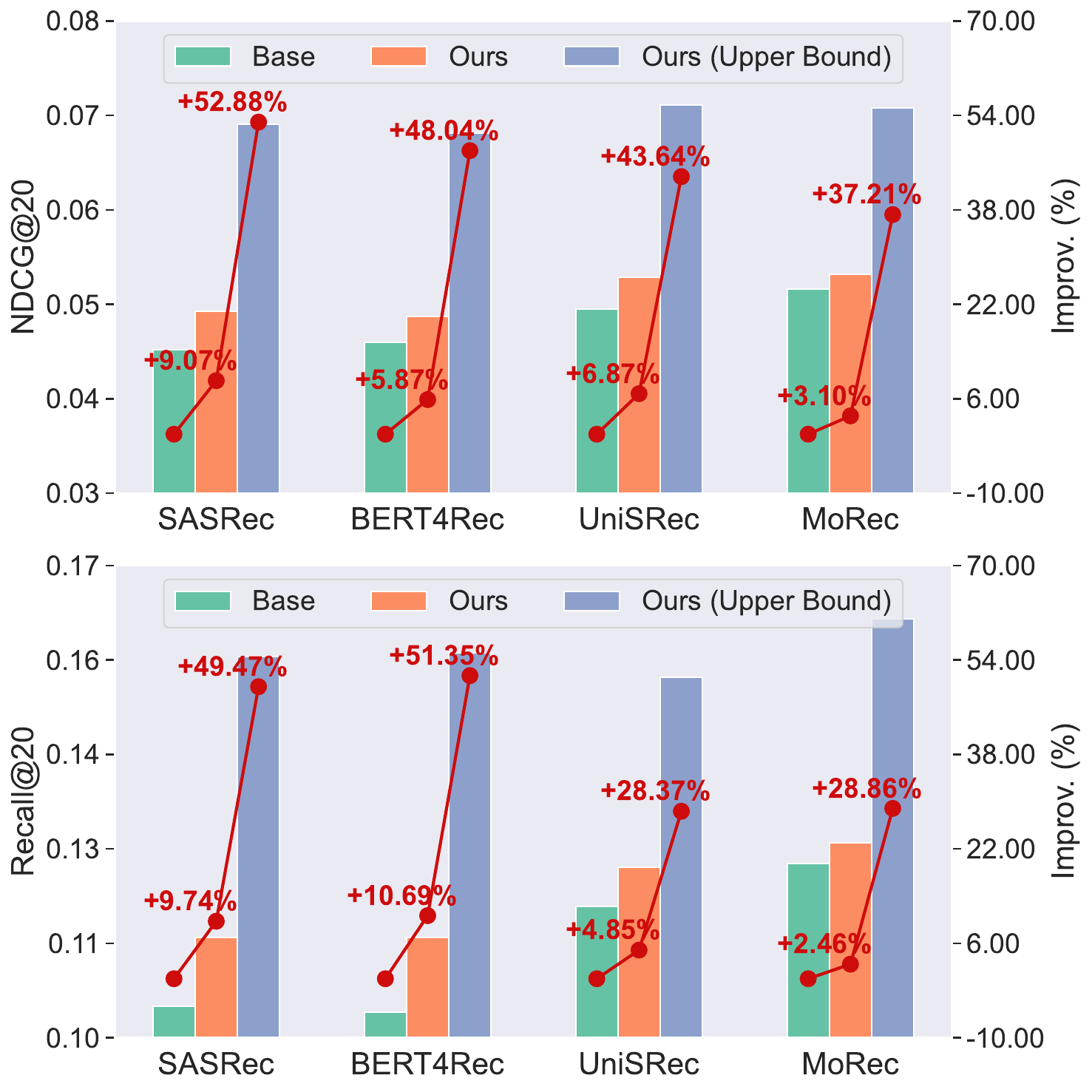}
  \vspace{-0.5cm}
  \caption{Empirical performance gains and potential upper bound analysis of optimal reasoning steps ($\mathbf{K=2}$) on Yelp dataset across different SeqRec models.}
  \label{fig:upper_bound}
  \vspace{-0.8cm}
\end{figure}

Our extensive experiments on five benchmark datasets demonstrate the effectiveness of the proposed ReaRec framework. In particular, the ReaRec achieves an average performance gain of 7.49\% across all metrics while incurring only 3.51\% additional inference latency (\emph{cf.} Sec.~\ref{sec:overall_performance} and Sec.~\ref{sec:reasoning_time}). 
Moreover, further analysis reveals several interesting empirical findings:
\textbf{(1) Enhancing modeling capability for underrepresented groups.}  The multi-step reasoning process steadily enhances the recommendation quality of users with sparse interactions and long-tail items.
\textbf{(2) Remarkable performance ceiling breakthrough.} Post-hoc optimal reasoning step analysis shows that our method elevates the performance ceilings for different backbone models by approximately 30\%-50\% (as shown in Fig.~\ref{fig:upper_bound}), highlighting its promising capability. We are optimistic that our proposed RecRec will open new avenues for exploring inference-time scaling for recommender systems. 

Our main contributions are summarized as follows:
\begin{itemize}[leftmargin=*]
  \item We propose \textbf{ReaRec}, a novel reasoning-enhanced sequential recommendation framework that empowers SeqRec models to perform implicit multi-step reasoning during inference. To our knowledge, this is the first work to systematically explore inference-time computational power within recommender systems.
  \item We introduce two reasoning learning strategies, ERL and PRL, which leverage the ideas of ensemble learning and curriculum learning to efficiently optimize the implicit reasoning process and alleviate reasoning degradation issues.
  \item Extensive experiments on five real-world datasets and various representative SeqRec models validate the generality and effectiveness of ReaRec. Notably, our detailed post-hoc analysis reveals that ReaRec can significantly raise the performance ceiling, achieving significant improvements by up to 50\%.
  \item We identify some challenges faced by current reasoning-enhanced recommendation methods and the future opportunities, stimulating a new research direction at the intersection of inference-time computing and sequential recommendation.
\end{itemize}

\section{Preliminary}
In this section, we formally define the sequential recommendation task and introduce the typical sequential recommendation pipeline.

\subsection{Problem Definition}
Formally, let $\mathcal{U}$ and $\mathcal{V}$ denote the sets of users and items, respectively, with $M = |\mathcal{U}|$ and $N = |\mathcal{V}|$ representing the number of users and items. For each user $u \in \mathcal{U}$, we define their chronological interaction sequence as $\mathcal{S}^u=[v_1^u, v_2^u, \ldots, v_{n_u}^u]$, where $n_u$ represents the length of the interaction sequence $\mathcal{S}_u$. Each item $v \in \mathcal{V}$ has a unique ID and a set of textual attributes (such as title, product feature, and other side information). These attributes are stored in a dictionary $\mathcal{D}_v = \{k_1:a_1, k_2:a_2,\ldots,k_m:a_m\}$, where $k_i$ and $a_i$ represent the key and value of the $i$-th attribute, respectively. Here, $m$ refers to the total number of attributes associated with item $v$. The overall text description for item $v$ is constructed by concatenating its attributes in the format of an unordered list: 
``The item information is as follows: \verb|\n|- $k_1$:$a_1$ \verb|\n|- $k_2$:$a_2$ \verb|\n| \ldots \verb|\n|- $k_m$:$a_m$''. 

The goal of sequential recommendation is to predict the next item a user will interact with, based on historical interaction data. 
Given the interaction sequences for all users $\mathcal{S}=\{\mathcal{S}^{u_1},\mathcal{S}^{u_2},\ldots,\mathcal{S}^{u_M}\}$, where $\mathcal{S}^{u_i}$ represents the interaction sequence of user $u_i$, and $\mathcal{S}^{u_i}_{1:t}=[v_1^u,v_2^u,\ldots,v_t^u]$ denotes the first $t$ interaction records of user $u_i$. 
Given the item embedding matrix $\mathbf{E} \in \mathbb{R}^{N\times d}$, where $d$ is the dimension of the item embedding, the sub-sequence $\mathcal{S}^{u_i}_{1:t}$ is encoded to obtain the corresponding item embeddings $\mathbf{E}^{u_i}_{1:t}=[\mathbf{e}_{v^u_1},\mathbf{e}_{v^u_2},\ldots,\mathbf{e}_{v^u_t}]$.
The recommender's learning objective is to maximize the prediction probability of the next item $v_{t+1}^{u_i}$ based on the historical interaction data, which is formally defined as
\begin{equation}
  \label{eq:learning_obj}
  \max_{\Theta} \sum_{u \in \mathcal{U}} \sum_{t=1}^{n_u-1} P(v_{t+1}^{u}|\mathcal{S}^{u}_{1:t};\Theta), 
\end{equation}
where $\Theta$ denotes the parameters of the recommendation model.

\subsection{Sequential Recommendation Pipeline}
In a typical sequential recommendation pipeline, users' historical interactions are first encoded into item embeddings. These item embeddings are then fed into a sequential model (\emph{e.g.}, transformer-based models) to produce a sequence representation, typically using the output from the final position (as illustrated in Fig.~\ref{fig:intro}(a)). Finally, this sequence representation is used to calculate similarity scores with candidate item embeddings (such as dot product~\cite{zhang2024uoep,zhang2024saqrec} or cosine similarity\cite{xu2025rallrec,li2023text}) to predict the probability of the user's interaction with the next item.

In general, mainstream sequential recommendation methods can be broadly categorized into two main types, distinguished primarily by their approaches to encoding item representations:

(1) \textbf{ID-based Encoding}: The ID-based approach uses one-hot encoding for the item's discrete representation and retrieves the item's embedding from the embedding matrix. Representative sequential recommendation methods employing this encoding approach include SASRec~\cite{kang2018self}, BERT4Rec\cite{sun2019bert4rec}, etc.

(2) \textbf{Text-based Encoding}: The text-based item representation usually involves feeding the item's string-formatted description into a pre-trained language model (such as BERT~\cite{reimers-gurevych-2019-sentence}, LLaMA~\cite{grattafiori2024llama}, etc.), and then utilizing average pooling or extracting hidden state from special positions (e.g., [CLS] or the last position) as the item's encoding~\cite{li2023text,liu2024large,geng2024breaking}. 
Popular recommendation models utilizing text-based encoding include UniSRec~\cite{hou2022towards}, MoRec~\cite{yuan2023go}, etc.

In this paper, since the proposed reasoning framework is model-agnostic, we omit the details of how item representations are obtained and consistently use $\mathbf{e}_v \in \mathbf{E}$ to denote item $v$' representations.

\section{Methodology}
In this section, we introduce ReaRec, a novel, simple, and highly scalable recommendation framework designed to unleash a model’s latent sequential reasoning capability. Instead of the traditional direct recommendation without reasoning, our approach leverages multi-step implicit reasoning to refine user representations, fully exploiting the computational potential of sequential models to approximate the true distribution of user interests.

\begin{figure}
  \centering
  \includegraphics[width=\linewidth]{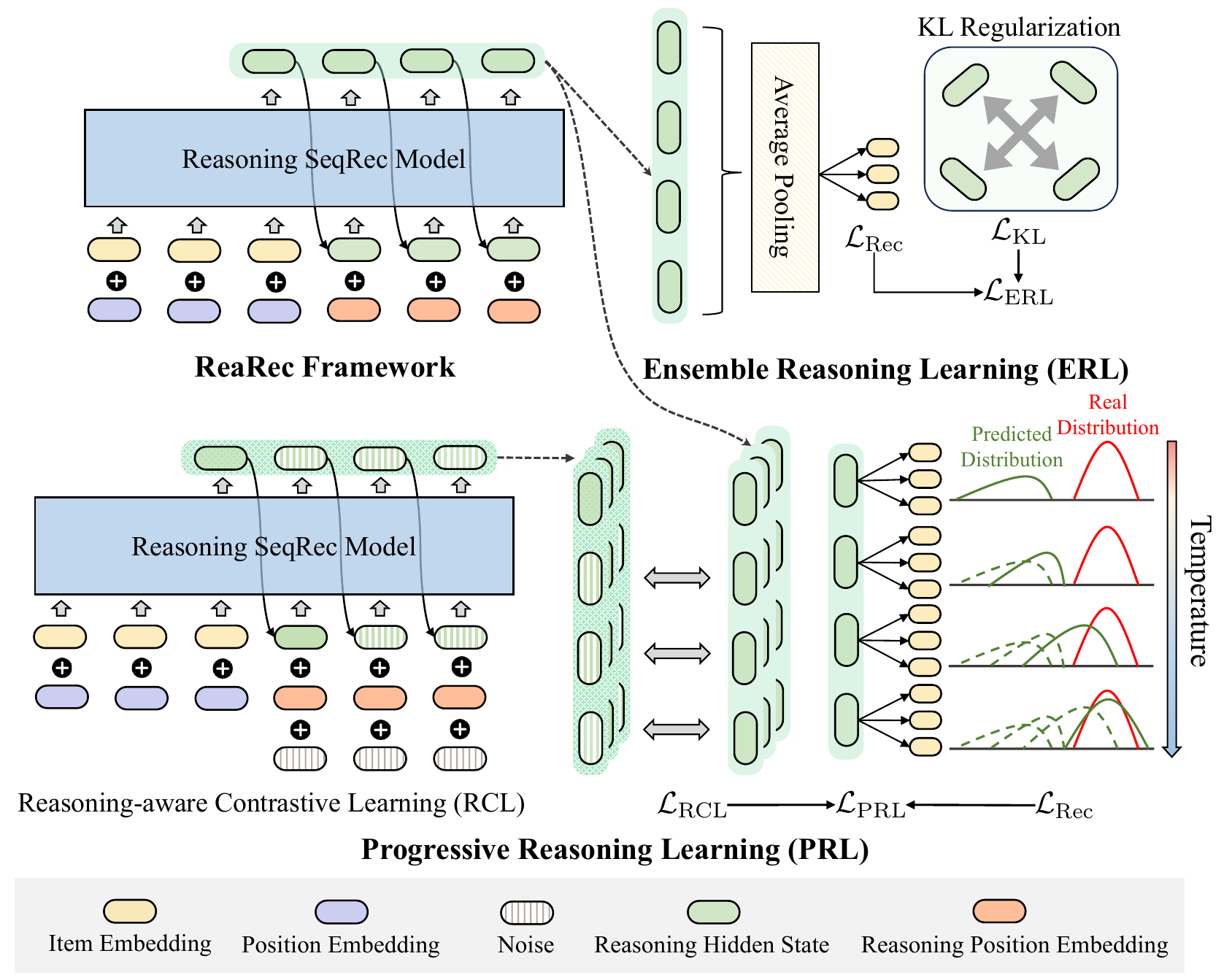}
    \caption{Overview of the proposed ReaRec framework and two reasoning-enhanced learning strategies: Ensemble Reasoning Learning and Progressive Reasoning Learning.}
  \label{fig:method}
\end{figure}

In what follows, we first introduce ReaRec, our foundational framework for inference-time computation extension (Sec.~\ref{sec:backbone}). We then propose two lightweight methods---Ensemble Reasoning Learning (Sec.~\ref{sec:erl}) and Progressive Reasoning Learning (Sec.~\ref{sec:prl})---to address the aforementioned challenges. Moreover, we give a deeper analysis of the proposed ReaRec framework (Sec.~\ref{sec:discussion}).
The overall framework of ReaRec is illustrated in Fig.~\ref{fig:method}.

\subsection{ReaRec Backbone}\label{sec:backbone}
Our proposed ReaRec is model-agnostic and can be easily integrated into a variety of sequential recommenders. To better explain our work, we illustrate our framework using the widely adopted transformer~\cite{vaswani2017attention} architecture in sequential recommendation tasks as an example, demonstrating how we extend computational capacity during inference with our backbone.

\subsubsection{\textbf{Self-attention Sequence Encoding}}
Given a user's historical sequence $\mathcal{S}_u=[v^u_1,v^u_2,\ldots,v^u_n]$, we can obtain the item embeddings of these $n$ items by looking up the embedding matrix $\mathbf{E}$.
To fully leverage sequential information, we inject \textit{Absolute Position Embeddings} into the item embeddings at the input layer. Specifically, for a given item $v$ at position $i$, the input representation is constructed by summing its item embedding $\mathbf{e}_{v}$ and the corresponding positional embedding $\mathbf{p}^I_{i}$:
\begin{equation}\label{eq:original_input}
  \mathbf{h}_{i}^{0} = \mathbf{e}_{v} + \mathbf{p}^I_{i},
\end{equation}
where $\mathbf{p}^I_{i}$ is obtained by looking up the learnable positional embedding matrix $\mathbf{P}^I\in \mathbb{R}^{n\times d}$.
Next, we develop the item sequence encoder $f(\cdot)$ by stacking multiple multi-head self-attention layers (denoted as $MHSA(\cdot)$) and point-wise feed-forward networks (denoted as $FFN(\cdot)$) to capture the complicated sequence features:
\begin{equation}
  \mathbf{H}^{l} = f(\mathbf{H}^{l-1}) = FFN(MHSA(\mathbf{H}^{l-1})),
\end{equation}
where $\mathbf{H}^l=[\mathbf{h}^l_1,\mathbf{h}^l_2,\ldots,\mathbf{h}^l_n]$ denotes the concatenated hidden states at the $l$-th layer.
In the conventional paradigm, the output at the last position of the final layer is directly used as the final user representation, \emph{i.e.}, $\mathbf{h}_u=\mathbf{H}^L[-1]$, where $L$ is the number of layers.

\subsubsection{\textbf{Extended Inference-Time Reasoning}}\label{sec:rearec}
Existing sequential recommenders that rely only on non-reasoning forward inference struggle to directly model item sequence patterns, fundamentally constrained by their limited computation power to capture nuanced user interest. To address this problem, we propose \textbf{implicit reasoning mechanism} to augment the computational capacity, enabling the enhanced refinement of user interest modeling to more precisely approximate real preference distributions.

Specifically, rather than directly using $\mathbf{H}^L[-1]$ as the user representation, we autoregressively feed the hidden state of the last position back into the encoder for \textbf{$\bm{K}$-pass forward computations}. By effectively increasing inference-time computation, this approach further unleashes the model's potential to capture intricate sequential dependencies.
However, this inference strategy deviates from the original objective of sequential recommendation models, namely next-item prediction.
To bridge this task gap, we introduce the \textbf{\textit{Reasoning Position Embedding (RPE)}}, denoted as $\mathbf{P}^{R}\in \mathbb{R}^{K\times d}$, to distinguish between the sequence encoding phase and the reasoning phase.
At the $k$-th reasoning step, the model's input embedding is defined as $\mathbf{H}^0\in \mathbb{R}^{(n+k-1)\times d}$. The first $n$ positions remain unchanged from the original input (\emph{i.e.}, Eq.~\eqref{eq:original_input}), while the latent representation $\mathbf{h}^0_{n+i}$ at position $n+i$ is calculated as the summation of the last output $\mathbf{h}^L_{n+i-1}$ from the previous step and the $i$-th reasoning position embedding $\mathbf{p}^R_{i}$:
\begin{equation}
  \mathbf{h}^0_{n+i} = \mathbf{h}^{L}_{n+i-1} + \mathbf{p}^R_{i}.
\end{equation}

To differentiate between item encoding outputs and reasoning outputs, we denote the hidden states of the model's final layer from position $n$ to $n+k$ as $\mathbf{R}=[\mathbf{r}_0, \mathbf{r}_1, \ldots, \mathbf{r}_{k}]$, where $\mathbf{r}_i\in\mathbb{R}^d$ represents the reasoning hidden state at the $i$-th step.
To obtain the user representation, a straightforward approach is to follow the traditional paradigm \emph{i.e.}, use the last reasoning output $\mathbf{r}_K$ as $\mathbf{h}_u$. 
Then, we calculate the predicted probability for the user $u$ as 
\begin{equation}\label{eq:hat_y}
    \hat{y} = \text{softmax}(\mathbf{h}_u\cdot \mathbf{E}^\top)
\end{equation}
and use cross-entropy loss as the recommendation objective:
\begin{equation}\label{base_rec_loss}
    \mathcal{L}_{\text{Rec}} = - \log \hat{y}_{v^+},
\end{equation}
where $\hat{y}_{v^+}$ denotes the prediction probability of the ground-truth item $v^+$ for user $u$'s next interaction.

However, this naive optimization objective still faces a critical issue: the lack of supervision signals for intermediate reasoning states makes the model vulnerable to the risk of reasoning pattern degradation.
Next, we introduce two simple yet effective reasoning learning strategies to address these challenges.

\subsection{Ensemble Reasoning Learning (ERL)}\label{sec:erl}
To provide effective supervised signals for the implicit reasoning process, we propose an \textit{Ensemble Reasoning Learning (ERL)} method. This approach uses the hidden states of different reasoning steps as multi-view representations of the user's evolving interests. In other words, we apply the idea of \textit{ensemble learning}~\cite{dong2020survey,sagi2018ensemble} to aggregate diverse reasoning results from different reasoning steps, thereby avoiding suboptimal performance caused by the final output alone. 

\subsubsection{\textbf{Multi-Step Reasoning Supervision}}
Specifically, we treat the reasoning hidden states from multiple steps as multi-vector user representations and apply cross-entropy loss (\emph{cf.} Eq.~\eqref{base_rec_loss}) to the ensembled sequence representation to enhance process guidance. Therefore, instead of using only the reasoning state from the last step, ERL employs an average pooling layer to aggregate reasoning hidden states across all steps to obtain the final user representation as $\mathbf{h}_u = \frac{1}{K}\sum_{i=0}^{K}\mathbf{r}_i$. The output distribution $\hat{y}$ is then computed following Eq.~\eqref{eq:hat_y}.

\subsubsection{\textbf{KL Divergence Regularization}}\label{sec:kl}
However, simply using the above recommendation objective for model training is obviously inefficient. The recommender may take shortcuts by directly copying the previous reasoning output to optimize the parameters, which can lead to a \textit{pattern collpase effect}, consequently undermining the advantage of computational scaling during inference processes.
To this end, inspired by the works~\cite{hwang2021multi,jin2024hgdl}, we introduce a Kullback-Leibler (KL) divergence constraint, a popular and simple regularization technique to mitigate the homogenization output issue. 
To be specific, we aim to increase the reasoning output diversity across different steps, encouraging the model's multi-step reasoning process to gather multi-view insights, and better model the user's complex interest distribution, ultimately contributing to the overall sequence recommendation performance.
Formally, we pair the predictive probability distributions of different reasoning states in pairwise combinations and maximize the KL divergence between these distribution pairs, which is equivalent to minimizing the following regularization term:
\begin{equation}
  \mathcal{L}_{\text{KL}} = -\sum_{i=0}^{K-1}\sum_{j=i+1}^{K} \text{KL}(\hat{y}^{(i)}\Vert\hat{y}^{(j)}).
\end{equation}
where $\hat{y}^{(i)}$ represents the logit output of the $i$-th reasoning step.

By combining the recommendation loss and the above KL regularization term, the overall learning objective for the ERL method is to minimize the following loss function:
\begin{equation}
  \mathcal{L}_{\text{ERL}} = \mathcal{L}_{\text{Rec}} + \lambda \mathcal{L}_{\text{KL}},
\end{equation}
where $\lambda$ is a hyperparameter that balances the constraint strength.

\subsubsection{\textbf{Inference Phase}}\label{sec:erl_inference}
In the inference phase, we apply an average pooling layer to aggregate the reasoning hidden states from all steps into the final user representation, \emph{i.e.}, $\mathbf{h}_u = \frac{1}{K}\sum_{i=0}^{K}\mathbf{r}_i$.
Then, we compute the inner product or cosine similarity (depending on the specific sequential recommendation algorithm) between user representation and all candidate item representations, with top-scoring items selected as the final recommendation list.

\subsection{Progressive Reasoning Learning (PRL)}\label{sec:prl}
Unlike the ensemble reasoning learning method, we explore another \textit{Progressive Reasoning Learning (PRL)} mechanism. The core idea is to design a progressive distribution sharpening strategy to guide the intermediate reasoning chains, gradually approximating the user's true preference distribution.
Intuitively, as the computational power allocated to the inference time increases, the recommendation model should be able to more accurately capture the fine-grained sequential features, narrowing the discrepancy between the predicted and actual user interest distribution.

\subsubsection{\textbf{Progressive Temperature Annealing (PTA)}}
Drawing an analogy the human cognitive process, as the thinking depth increases, reasoning pathways become progressively refined until converging toward optimal solutions.
Similarly, we expect that as the model's computations increases, the recommender would gradually clarify the user's interest evolving patterns, which is manifested as sharper predicted distributions.
Inspired by this motivation, we propose a simple \textit{Progressive Temperature Annealing (PTA)} method to guide the reasoning process.
To achieve this, we first introduce a temperature coefficient, $\tau_k$, for the $k$-th reasoning step to adjust the predicted distribution sharpness, which is formulated as follows:
\begin{equation}
  \begin{gathered}
    \tau_k = \tau * \alpha^{K-k},\\
    \hat{y}^{(k)} = \text{softmax}(\mathbf{r}_k\cdot \mathbf{E}^\top / \tau_k),
  \end{gathered}
  \end{equation}
where $\tau$ is the base temperature, and $\alpha$ is a hyperparameter that controls the temperature decay rate.

In contrast to ensemble reasoning learning method, we apply separate recommendation losses to each reasoning hidden state to inject process supervision into the reasoning process, as follows:
\begin{equation}\label{eq:prl_rec_loss}
  \mathcal{L}_{\text{Rec}} = - \sum_{k=0}^{K} \log \hat{y}_{v^+}^{(k)},
\end{equation}
where $\hat{y}_{v^+}^{(k)}$ represents the logit corresponding to the $v^+$ item.
With this lean annealing strategy, the model is encouraged to explore a broader solution space in the early reasoning stage, preventing it from getting stuck in local optima. Then, as the reasoning process progresses, the value of $\tau_k$ is gradually reduced to narrow the search space, guiding the model towards the global optimum.
Thus, the proposed PTA can more effectively approximate the user's true preference distribution.

\subsubsection{\textbf{Reasoning-aware Contrastive Learning (RCL)}}\label{sec:rcl}
However, relying solely on the temperature annealing strategy may not be sufficient to support the generalization ability of progressive reasoning learning. This is because, during the reasoning process, the model may suffer from the \textit{reasoning bias}, where the model's reasoning direction deviates from the correct user interest distribution, ultimately leading to the accumulation of reasoning errors and deteriorating the reasoning capability.
To address the above challenge, we design a novel \textit{Reasoning-aware Contrastive Learning (RCL)} method to enhance the model's robust reasoning ability.

Specifically, we simulate the preceding accumulated reasoning error by injecting noise vectors into the reasoning states for each step, producing the noised reasoning input as follows:
\begin{equation}
  \mathbf{\tilde{h}}^0_{n+i} = \mathbf{h}_{n+i}^0 + \bm{\epsilon}, \quad i \in \{1,2,\ldots,K\},
\end{equation}
where $\mathbf{h}_{n+i}^0$ is defined according to Eq.~\eqref{eq:original_input}. The vector $\bm{\epsilon}$ represents the added noise embedding, sampled from a normal distribution, \emph{i.e.}, $\bm{\epsilon} \sim \mathcal{N}(\mathbf{0}, \gamma\mathbf{I})$, where $\mathbf{I} \in \mathbb{R}^d$ is the identity matrix of dimension $d$ and $\gamma$ controls the noise intensity.
Then, we can obtain the new hidden state view $\mathbf{\tilde{R}}=[\mathbf{\tilde{r}}_1,\mathbf{\tilde{r}}_2,\ldots,\mathbf{\tilde{r}}_K]$ by feeding the noised input into the transformer encoder.

To enhance the model's robustness in reasoning denoising, we design a self-supervised task based on \textit{Mutual Information Maximization (MIM)}~\cite{tschannen2019mutual,viola1997alignment}. Formally, given variables $X$ and $Y$, the \textit{Mutual Information (MI)} measures the reduction in uncertainty of X after observing Y, which is defined as:
\begin{equation*}\label{eq:mi}
  I(X,Y) = H(X) - H(X|Y),
\end{equation*}
where $H(\cdot)$ and $H(\cdot|\cdot)$ denote the entropy and conditional entropy of the random variable, respectively.
By maximizing the MI between the original hidden states $\mathbf{R}$ and the denoised hidden states $\mathbf{\tilde{R}}$, it can effectively force the model to capture the essential sequential information from the user behavior data and historical reasoning process, achieving \textbf{self-reflection in the implicit thought space}.

However, directly maximizing mutual information is not feasible due to the intractability of the high-dimensional probability distribution estimation.
Inspired by recent works~\cite{tian2020makes,wu2023understanding}, we propose an InfoNCE-based reasoning contrastive learning method to optimize the lower bound of mutual information, which is defined as:
\begin{equation}
  \begin{gathered}
    \mathcal{L}_{\text{RCL}} = -\sum_{k=1}^{K} \log \frac{\exp(\text{sim}(\mathbf{\tilde{r}}_k, \mathbf{r}_k^+)/\tau)}{\exp(\text{sim}(\mathbf{\tilde{r}}_k, \mathbf{r}_k^+)/\tau) + \sum_{\mathbf{r}_k^- \in \mathbf{R}_k^-} \exp(\text{sim}(\mathbf{\tilde{r}}_k, \mathbf{r}_k^-)/\tau)},
  \end{gathered}
\end{equation}
where $\text{sim}(\cdot)$ denotes the dot product similarity function, $\mathbf{r}_k^+$ and $\mathbf{r}_k^-$ indicate the positive and negative contrastive hidden states at the $k$-th step, respectively. For the negative sample set $\mathbf{R}_k^-$, analogous to existing methods~\cite{tang2024towards,yu2022graph}, we utilize the $k$-th step reasoning states corresponding to the other item sequences within the same batch.

By combining the recommendation loss and the reasoning contrastive loss, we can derive the overall objective function for the PRL method as follows:
\begin{equation}
  \mathcal{L}_{\text{PRL}} = \mathcal{L}_{\text{Rec}} + \mathcal{L}_{\text{RCL}}.
\end{equation}

\subsubsection{\textbf{Inference Phase}}
During inference, we directly adopt the final reasoning step's output as the user representation, \emph{i.e.}, $\mathbf{h}_u = \mathbf{r}_K$. Then, similar to Sec.~\ref{sec:erl_inference}, we compute similarity scores between $\mathbf{h}_u$ and the candidate item embedding matrix $\mathbf{E}$ to generate the recommendation list for the user $u$.

\subsection{Discussion}\label{sec:discussion}
\subsubsection{\textbf{Principle Analysis}}
The ReaRec framework fundamentally extends the model's modeling capability by strategically increasing inference-time computational amounts. By autoregressively feeding the reasoning hidden states into the sequence encoder, the model continuously \textbf{deepens feature crossing depth}, capturing finer-grained sequence characteristics and eventually improving the recommendation performance.
Moreover, the proposed ERL and PRL methods unleash the latent reasoning power of sequential recommenders in different ways. The ERL integrates multi-level deep crossing features into the final user representation, while the latter, based on the concept of curriculum learning, gradually uncovers more complex intent evolution patterns as the reasoning process progresses, moving closer to real user interest distribution.

\subsubsection{\textbf{Time and Space Complexity}}
In this part, we provide a detailed analysis of the time and space complexity of the proposed ReaRec framework as follows:
\begin{itemize}[leftmargin=*]
  \item \textbf{Time Complexity}. Suppose the user sequence length is $C$, we first analyze the base backbone time without reasoning extension. The input sequence passes through $L$ layers of $MHSA$ modules ($O(C^2d+Cd^2)$) and $FFN$ modules ($O(Cd^2)$), resulting in a total time complexity of $O(L(C^2d+Cd^2))$.
  For the reasoning-enhanced phase, we employ \textbf{\textit{KV Caching}} technique to store history key-value pairs, eliminating redundant computations. Specifically, at the $k$-th reasoning step, the time complexity of the $MHSA$ and $FFN$ are $O((C+k-1)d)$ and $O(d^2)$, respectively. After applying $L$ transformer blocks and $K$ steps of reasoning, the total additional time complexity overhead is $O(L(K(C+K)d + Kd^2))$. Since the number of reasoning steps (\emph{e.g.}, $K=2$) is usually much smaller than $C$, this overhead is simplified to $O(L(KCd+Kd^2))$.
  Therefore, our framework does not bring significant time cost, making it suitable for practical deployment in real-world industry recommender systems.
  \item \textbf{Space Complexity}. Our method only adds $K$ $d$-dimensional reasoning position embeddings $\mathbf{P}^R$, which is almost negligible compared to the original model parameters. Therefore, our framework is highly lightweight and flexible. 
\end{itemize}

\section{Experiments}
{
\renewcommand{\arraystretch}{1.15}
\begin{table}[t]
  \centering
  \caption{The statistics of experimental datasets.}
  \resizebox{\linewidth}{!}{
  \begin{tabular}{*{7}{c}}
    \toprule
    Dataset & Yelp & \makecell{Video \&\\ Games} & Software & \makecell{CDs \&\\ Vinyl} & \makecell{Baby \&\\ Products} \\
    \midrule
    \#Users & 13,083 & 89,021 & 30,049 & 35,238 & 140,292 \\
    \#Items & 10,697 & 22,933 & 16,705 & 87,969 & 30,689 \\
    \#Avg. Inter. / User & 33.92 & 5.96 & 5.59 & 14.59 & 5.57 \\
    \#Avg. Inter. / Item & 41.49 & 23.15 & 10.06 & 5.84 & 25.44 \\
    \#Avg. Inter. & 443,807 & 530,989 & 168,029 & 513,991 & 780,809 \\
    Sparisty & 99.68\% & 99.97\% & 99.97\% & 99.98\% & 99.98\% \\
    \bottomrule
\end{tabular}

  }
  \label{tab:dataset}
\end{table}
}

In this section, we conduct extensive experiments and analyses to demonstrate the superiority of our proposed ReaRec framework.

\subsection{Experimental Setup}
\subsubsection{\textbf{Datasets}}
To evaluate the effectiveness of our proposed methods, we conduct extensive experiments on five real-world recommendation datasets from Yelp and Amazon platforms. The detailed statistics of the datasets are summarized in Table~\ref{tab:dataset}.

(1) \textbf{Yelp}\footnote{\url{https://business.yelp.com/data/resources/open-dataset/}}: This dataset originates from a well-known business review website, providing rich multidimensional data support for studying user behaviors and business attributes. We treat interactions with ratings greater than 3 as positive samples and apply 20-core filtering to preprocess the data. For textual encoding, we retain the name, location (city and state), and business categories as item information. The dataset is chronologically split into training, validation, and test sets based on two timestamp thresholds: September 4, 2018 and May 12, 2020.

(2) \textbf{Amazon 2023}\footnote{\url{https://amazon-reviews-2023.github.io/}}: This dataset is derived from Amazon, a leading global e-commerce platform. We select datasets from four domains: Video \& Games, Software, CDs \& Vinyl, and Baby \& Products. For textual features, we retain the product attributes like title, description, and price. Similarly, we treat user-item interactions with user ratings greater than 3 as positive samples. To ensure data quality, we filter out users with fewer than 5 interactions for Video \& Games, Software, Baby \& Products, and fewer than 10 interactions for CDs \& Vinyl. For dataset splitting, we follow the official absolute timestamps to partition item sequences\footnote{\url{https://amazon-reviews-2023.github.io/data_processing/5core.html}}. This aligns well with real-world scenarios and facilitates fair performance comparisons within the recommendation research community.

\subsubsection{\textbf{Evaluation Metrics}}
We adopt top-k \textit{Normalized Discounted Cumulative Gain (NDCG)} and top-k \textit{Recall} to measure the recommendation performance, which are widely used in related sequential recommendation research~\cite{yang2023debiased,tan2021sparse,chen2018sequential}. In this paper, we specifically report \textbf{NDCG@\{10,20\}}, which assesses both the relevance and ranking quality of the top-k recommended items, and \textbf{Recall@\{10,20\}}, which evaluates the ability of the model to recall the ground-truth items in the top-k list.

{
\renewcommand{\arraystretch}{1.15}
\begin{table*}[htbp]
  \centering
  \caption{Performance comparison of different ID-based models on five datasets. `N' and `R' indicate NDCG and Recall metrics, respectively. `Avg.' represents the average improvement rate across all metrics (\emph{i.e.}, NDCG@\{10,20\} and Recall@\{10,20\}). Performance improvements are indicated by ``$\uparrow$'', while performance declines are indicated by ``$\downarrow$''.}
  \resizebox{\textwidth}{!}{
  \begin{tabular}{
    c c 
    c c c c c @{\hspace{0.4em}} 
    c @{\hspace{0.4em}}
    c c c c c @{\hspace{0.4em}}
}
    \toprule
    \multirow{2}{*}{Dataset} & \multirow{2}{*}{Method} & \multicolumn{5}{c}{SASRec} & & \multicolumn{5}{c}{BERT4Rec} \\
    \cmidrule{3-7} \cmidrule{9-13}
    & & N@10 & N@20 & R@10 & R@20 & Avg. & & N@10 & N@20 & R@10 & R@20 & Avg. \\
    \midrule
    \multirow{5}{*}{Yelp} & Base & 0.0347 & 0.0452 & 0.0626 & 0.1047 & -  &  & 0.0364          & 0.046  & 0.0653 & 0.1038 &  -      \\
    & \makecell{+ERL\\(Improv.)} & \upcell{0.0383}{10.37} & \upcell{0.0474}{4.87} & \upcell{0.0691}{10.38} & \upcell{0.1056}{0.86} & \upwardbf{6.62\%}  &  & \upcell{0.0371}{1.92} & \upcell{0.0476}{3.48} & \upcell{0.0661}{1.23} & \upcell{0.1077}{3.76} & \upwardbf{2.60\%} \\
    & \makecell{+PRL\\(Improv.)} & \upcell{0.0388}{11.82} & \upcell{0.0493}{9.07} & \upcell{0.073}{16.61}  & \upcell{0.1149}{9.74} & \upwardbf{11.81\%} &  & \upcell{0.0377}{3.57} & \upcell{0.0487}{5.87} & \upcell{0.0708}{8.42} & \upcell{0.1149}{10.69} & \upwardbf{7.14\%}  \\
    \hline
    \multirow{5}{*}{Video \& Games} & Base & 0.0284 & 0.0353 & 0.0542 & 0.0816 & -      &  & 0.0289 & 0.0355 & 0.0548 & 0.0810 & -      \\
    & \makecell{+ERL\\(Improv.)} & \upcell{0.0301}{5.99} & \upcell{0.0385}{9.07} & \upcell{0.0581}{7.20} & \upcell{0.0915}{12.13} & \upwardbf{8.59\%} &  & \upcell{0.0311}{7.61} & \upcell{0.0375}{5.63} & \upcell{0.0578}{5.47} & \upcell{0.0832}{2.72} & \upwardbf{5.36\%} \\
    & \makecell{+PRL\\(Improv.)} & \upcell{0.0299}{5.28} & \upcell{0.0379}{7.37} & \upcell{0.0572}{5.54} & \upcell{0.0890}{9.07} & \upwardbf{6.81\%} &  & \upcell{0.0306}{5.88} & \upcell{0.0380}{7.04} & \upcell{0.0584}{6.57} & \upcell{0.0879}{8.52} & \upwardbf{7.00\%} \\
    \hline
    \multirow{5}{*}{Software} & Base & 0.0696 & 0.0895 & 0.1468 & 0.2264 & -      &  & 0.0710 & 0.0893 & 0.1530 & 0.2258 & -      \\
    & \makecell{+ERL\\(Improv.)} & \upcell{0.0743}{6.75} & \upcell{0.0935}{4.47} & \downcell{0.1456}{0.82} & \downcell{0.2224}{1.77} & \upwardbf{2.16\%} &  & \upcell{0.0769}{8.31} & \upcell{0.0964}{7.95} & \upcell{0.1554}{1.57} & \upcell{0.2328}{3.10} & \upwardbf{5.23\%} \\
    & \makecell{+PRL\\(Improv.)} & \upcell{0.0739}{6.18} & \upcell{0.0949}{6.03} & \upcell{0.1488}{1.36} & \upcell{0.2324}{2.65} & \upwardbf{4.06\%} &  & \upcell{0.0762}{7.32} & \upcell{0.0976}{9.29} & \downcell{0.1500}{1.96} & \upcell{0.2350}{4.07} & \upwardbf{4.68\%} \\
    \hline
    \multirow{5}{*}{CDs \& Vinyl} & Base & 0.0148 & 0.0174 & 0.0317 & 0.0419 & -       &  & 0.0149 & 0.0185 & 0.0326 & 0.0468 & -       \\
    & \makecell{+ERL\\(Improv.)} &  \upcell{0.0182}{22.97} & \upcell{0.0212}{21.84} & \upcell{0.0363}{14.51} & \upcell{0.0482}{15.04} & \upwardbf{18.59\%} &  & \upcell{0.0165}{10.74} & \upcell{0.0208}{12.43} & \upcell{0.0354}{8.59} & \upcell{0.0524}{11.97} & \upwardbf{10.93\%} \\
    & \makecell{+PRL\\(Improv.)} &  \upcell{0.0155}{4.73} & \upcell{0.0195}{12.07} & \downcell{0.0315}{0.63} & \upcell{0.0470}{12.17}  & \upwardbf{7.08\%}  &  & \upcell{0.0162}{8.72} & \upcell{0.0202}{9.19} & \upcell{0.0334}{2.45} & \upcell{0.0496}{5.98} & \upwardbf{6.59\%} \\
    \hline
    \multirow{5}{*}{Baby \& Products} & Base & 0.0112 & 0.0157 & 0.0260  & 0.0437 & -       &  & 0.0109 & 0.0154 & 0.0257 & 0.0439 & -       \\
    & \makecell{+ERL\\(Improv.)} & \upcell{0.0116}{3.57} & \upcell{0.0164}{4.46} & \downcell{0.0228}{12.31} & \downcell{0.0418}{4.35} & $\downarrow$2.16\% &  & \upcell{0.0148}{35.78} & \upcell{0.0195}{26.62} & \upcell{0.0293}{9.57} & \upcell{0.0481}{14.01} & \upwardbf{21.49\%} \\
    & \makecell{+PRL\\(Improv.)} & \upcell{0.0135}{20.54} & \upcell{0.0178}{13.38} & \upcell{0.0281}{8.08} & \upcell{0.0451}{3.20} & \upwardbf{11.30\%} &  & \upcell{0.0140}{28.44}  & \upcell{0.0185}{20.13} & \upcell{0.0291}{6.15} & \upcell{0.0466}{13.23} & \upwardbf{16.99\%} \\
    \bottomrule
\end{tabular}

  }
  \label{tab:id_main_exp}
\end{table*}
}

{
\renewcommand{\arraystretch}{1.15}
\begin{table*}[htbp]
  \centering
  \caption{Performance comparison of different Text-based models on five datasets. `N' and `R' indicate NDCG and Recall metrics, respectively. `Avg.' represents the average improvement rate across all metrics (\emph{i.e.}, NDCG@\{10,20\} and Recall@\{10,20\}). Performance improvements are indicated by ``$\uparrow$'', while performance declines are indicated by ``$\downarrow$''.}
  \resizebox{\textwidth}{!}{
  \begin{tabular}{
    c c 
    c c c c c @{\hspace{0.4em}} 
    c @{\hspace{0.4em}}
    c c c c c @{\hspace{0.4em}}
}
    \toprule
    \multirow{2}{*}{Dataset} & \multirow{2}{*}{Method} & \multicolumn{5}{c}{UniSRec} & & \multicolumn{5}{c}{MoRec} \\
    \cmidrule{3-7} \cmidrule{9-13}
    & & N@10 & N@20 & R@10 & R@20 & Avg. & & N@10 & N@20 & R@10 & R@20 & Avg. \\
    \midrule
    \multirow{5}{*}{Yelp} & Base & 0.0380 & 0.0495 & 0.0737 & 0.1195 & -      &  & 0.0391 & 0.0516 & 0.0757 & 0.1258 & -      \\
    & \makecell{+ERL\\(Improv.)} & \upcell{0.0406}{6.84} & \upcell{0.0521}{5.25} & \upcell{0.0770}{4.48} & \upcell{0.1227}{2.68} & \upwardbf{4.81\%} &  & \upcell{0.0417}{6.65} & \upcell{0.0531}{2.91} & \upcell{0.0832}{9.91} & \upcell{0.1283}{1.99} & \upwardbf{5.36\%} \\
    & \makecell{+PRL\\(Improv.)} & \upcell{0.0413}{8.68} & \upcell{0.0529}{6.87} & \upcell{0.0788}{6.92} & \upcell{0.1253}{4.85} & \upwardbf{6.83\%} &  & \upcell{0.0410}{4.86} & \upcell{0.0532}{3.10} & \upcell{0.0804}{6.21} & \upcell{0.1289}{2.46} & \upwardbf{4.16\%} \\
    \hline
    \multirow{5}{*}{Video \& Games} & Base & 0.0328 & 0.0421 & 0.0683 & 0.1054 & -       &  & 0.0350 & 0.0438 & 0.0716 & 0.1065 & -      \\
    & \makecell{+ERL\\(Improv.)} & \upcell{0.0364}{10.98} & \upcell{0.0440}{4.51} & \upcell{0.0711}{4.10} & \downcell{0.1015}{3.70} & \upwardbf{3.97\%}  &  & \upcell{0.0392}{12.00} & \upcell{0.0485}{10.73} & \upcell{0.0744}{3.91} & \upcell{0.1112}{4.41} & \upwardbf{7.76\%} \\
    & \makecell{+PRL\\(Improv.)} & \upcell{0.0352}{7.32} & \upcell{0.0433}{2.85} & \downcell{0.0658}{3.66} & \downcell{0.0982}{6.83} & $\downarrow$0.08\% &  & \upcell{0.0371}{6.00} & \upcell{0.0462}{5.48} & \downcell{0.0708}{1.12} & \upcell{0.1067}{0.19} & \upwardbf{2.64\%} \\
    \hline
    \multirow{5}{*}{Software} & Base & 0.0820 & 0.1041 & 0.1643 & 0.2522 & -      &  & 0.0846 & 0.1050 & 0.1697 & 0.2510 & -      \\
    & \makecell{+ERL\\(Improv.)} & \upcell{0.0851}{3.78} & \upcell{0.1075}{3.27} & \upcell{0.1669}{1.58} & \upcell{0.2556}{1.35} & \upwardbf{2.49\%} &  & \upcell{0.0881}{4.14} & \upcell{0.1071}{2.00} & \upcell{0.1711}{0.82} & \downcell{0.2466}{1.75} & \upwardbf{1.30\%} \\
    & \makecell{+PRL\\(Improv.)} & \upcell{0.0869}{5.98} & \upcell{0.1076}{3.36} & \upcell{0.1687}{2.68} & \downcell{0.2518}{0.16} & \upwardbf{2.96\%} &  & \upcell{0.0917}{8.39} & \upcell{0.1120}{6.67} & \upcell{0.1723}{1.53} & \upcell{0.2532}{0.88} & \upwardbf{4.37\%} \\
    \hline
    \multirow{5}{*}{CDs \& Vinyl} & Base & 0.0150 & 0.0208 & 0.0298 & 0.0527 & -       &  & 0.0186 & 0.0235 & 0.0405 & 0.0604 & -      \\
    & \makecell{+ERL\\(Improv.)} &  \upcell{0.0208}{38.67} & \upcell{0.0259}{24.52} & \upcell{0.0428}{43.62} & \upcell{0.0629}{19.35} & \upwardbf{31.54\%} &  & \upcell{0.0199}{6.99} & \upcell{0.0248}{5.53} & \upcell{0.0417}{2.96} & \upcell{0.0609}{0.83} & \upwardbf{4.08\%} \\
    & \makecell{+PRL\\(Improv.)} &  \upcell{0.0191}{27.33} & \upcell{0.0253}{21.63} & \upcell{0.0394}{32.21} & \upcell{0.0640}{21.44} & \upwardbf{25.66\%} &  & \upcell{0.0198}{6.45} & \upcell{0.0249}{5.96} & \upcell{0.0417}{2.96} & \upcell{0.0618}{2.32} & \upwardbf{4.42\%} \\
    \hline
    \multirow{5}{*}{Baby \& Products} & Base & 0.0152 & 0.0199 & 0.0315 & 0.0501 & -       &  & 0.0176 & 0.0231 & 0.0371 & 0.0588 & -      \\
    & \makecell{+ERL\\(Improv.)} & \upcell{0.0183}{20.39} & \upcell{0.0239}{20.10} & \upcell{0.0367}{16.51} & \upcell{0.0589}{17.56} & \upwardbf{18.64\%} &  & \upcell{0.0184}{4.55} & \upcell{0.0242}{4.76} & \upcell{0.0373}{0.54} & \upcell{0.0602}{2.38} & \upwardbf{3.06\%} \\
    & \makecell{+PRL\\(Improv.)} & \upcell{0.0182}{19.74} & \upcell{0.0236}{18.59} & \upcell{0.0359}{13.97} & \upcell{0.0575}{14.77} & \upwardbf{16.77\%} &  & \upcell{0.0189}{7.39} & \upcell{0.0247}{6.93} & \upcell{0.0376}{1.35} & \upcell{0.0611}{3.91} & \upwardbf{4.89\%} \\
    \bottomrule
\end{tabular}

  }
  \label{tab:text_main_exp}
\end{table*}
}

\subsubsection{\textbf{Baselines}}
To thoroughly evaluate the generality of our proposed reasoning-enhanced framework, we conduct comprehensive benchmarking across different types of sequential recommendation models, including both ID-based and text-based encoding methods. The baselines are as follows:
For \underline{ID-based encoding methods}, we compare our methods with the following state-of-the-art models: 
\begin{itemize}
  \item \textbf{SASRec}~\cite{kang2018self}, a representative baseline for sequential recommendation, employs causal multi-head attention mechanism to capture sequential patterns in user interaction data.
  \item \textbf{BERT4Rec}~\cite{sun2019bert4rec}, a widely-used sequential model, leverages bidirectional self-attention layers for deeper contextual information infusion across user behavior sequences.
\end{itemize}
For \underline{Text-based encoding methods}, we adopt the following algorithms as backbones:
\begin{itemize}
  \item \textbf{UniSRec}~\cite{hou2022towards} utilizes parameter whitening and a \textit{Mixture-of-Experts (MoE)} adaptor to learn universal item and sequence representations from textual features, which effectively addresses cold-start and data sparsity challenges.
  \item \textbf{MoRec}~\cite{yuan2023go} replaces traditional ID features by incorporating advanced text and visual encoders (\emph{e.g.}, RoBERTa~\cite{liu2019roberta} and ViT~\cite{dosovitskiy2020image}) to model the multimodal representations of items. 
\end{itemize}

\subsubsection{\textbf{Implementation Details}}
We conduct all experiments on 8 NVIDIA A100 GPUs. To ensure a fair comparison, we set the embedding size and batch size for all methods to 256 and 2048, respectively. We optimize all models using the Adam~\cite{kingma2014adam} optimizer with a learning rate of 0.001 and follow previous work~\cite{sun2019bert4rec} by adopting GeLU as the activation function.
Following the existing works~\cite{xie2022contrastive,chen2022intent}, we truncate user sequences to a maximum length of 50 across all datasets. Since our framework is model-agnostic, it can be seamlessly integrated into various sequential recommendation models. In particular, for BERT4Rec's bidirectional Transformer, we employ a \textit{Prefix Masking} strategy, where the item sequence part utilizes bidirectional attention, while the reasoning adopts unidirectional attention. Early stopping is triggered if the metrics on the validation set do not improve over 10 consecutive epochs.
For item-based methods, we follow previous work~\cite{liu2024large} by using \textit{LLaMA-3.1-8B}~\cite{grattafiori2024llama} to encode item textual features. In particular, we apply \textit{Principle Component Analysis (PCA)} to the averaged hidden states from the last layer, preserving core features and distilling 768-dimensional model representations.
For ERL method, we search for the KL regularization hyperparameter $\lambda$ within $\{0.001, 0.005, 0.01, 0.05, 0.1\}$.
For PRL method, we set the noise strength $\gamma=0.01$ and tune the base temperature $\tau$ and temperature decay rate $\alpha$ over the ranges $\{0.05, 0.1, 0.5, 1.0, 2.0, 5.0\}$ and $\{1.0, 1.2, 1.5, 2.0, 5.0, 10.0\}$, respectively. Our codes will be available at \textcolor{blue}{\url{https://github.com/TangJiakai/ReaRec}}.

\subsection{Overall Performance}\label{sec:overall_performance}
The recommendation performance of ID-based and text-based sequential models across all datasets is summarized in Table~\ref{tab:id_main_exp} and Table~\ref{tab:text_main_exp}, respectively. We derive the following observations:
\begin{itemize}[leftmargin=*]
  \item For ID-based recommenders (\emph{i.e.}, SASRec and BERT4Rec), we can find that BERT4Rec slightly outperforms SASRec at different metrics on most datasets. This suggests that incorporating both left and right contextual information enhances the model's ability to capture sequential patterns more effectively.
  \item Text-based methods (\emph{i.e.}, UniSRec and MoRec) consistently outperform ID-based models across all datasets. For instance, on the Yelp dataset, UniSRec achieves a 9.51\% improvement in NDCG@20 and a 14.14\% increase in Recall@20 compared to SASRec. This improvement can be attributed to the ability of text-based models to leverage powerful language models for encoding item information, effectively mitigating data sparsity issues. In other words, by learning domain-invariant representations from textual feature spaces, these approaches effectively alleviate the recommendation bias, where underrepresented users and items are dominated by popular ones.
  \item Our proposed ERL and PRL methods, based on the ReaRec framework, consistently and significantly surpass baseline models at most cases. For example, for ID-based methods, ERL and PRL built on SASRec achieve average improvements of 6.76\% and 8.21\% respectively across all metrics on five datasets. Similarly, for text-based methods, ERL and PRL built on UniSRec outperform the base model by 12.29\% and 10.43\% on average. Unlike conventional SeqRec models, our reasoning-enhanced framework employs latent-space computations during the inference phase to deepen the feature crossing depth. This effectively unlock the latent reasoning power of various SeqRec backbones, demonstrating that increasing inference-time computation is a promising avenue for improving recommendation performance.
\end{itemize}

\begin{figure}
  \centering
  \includegraphics[width=\linewidth]{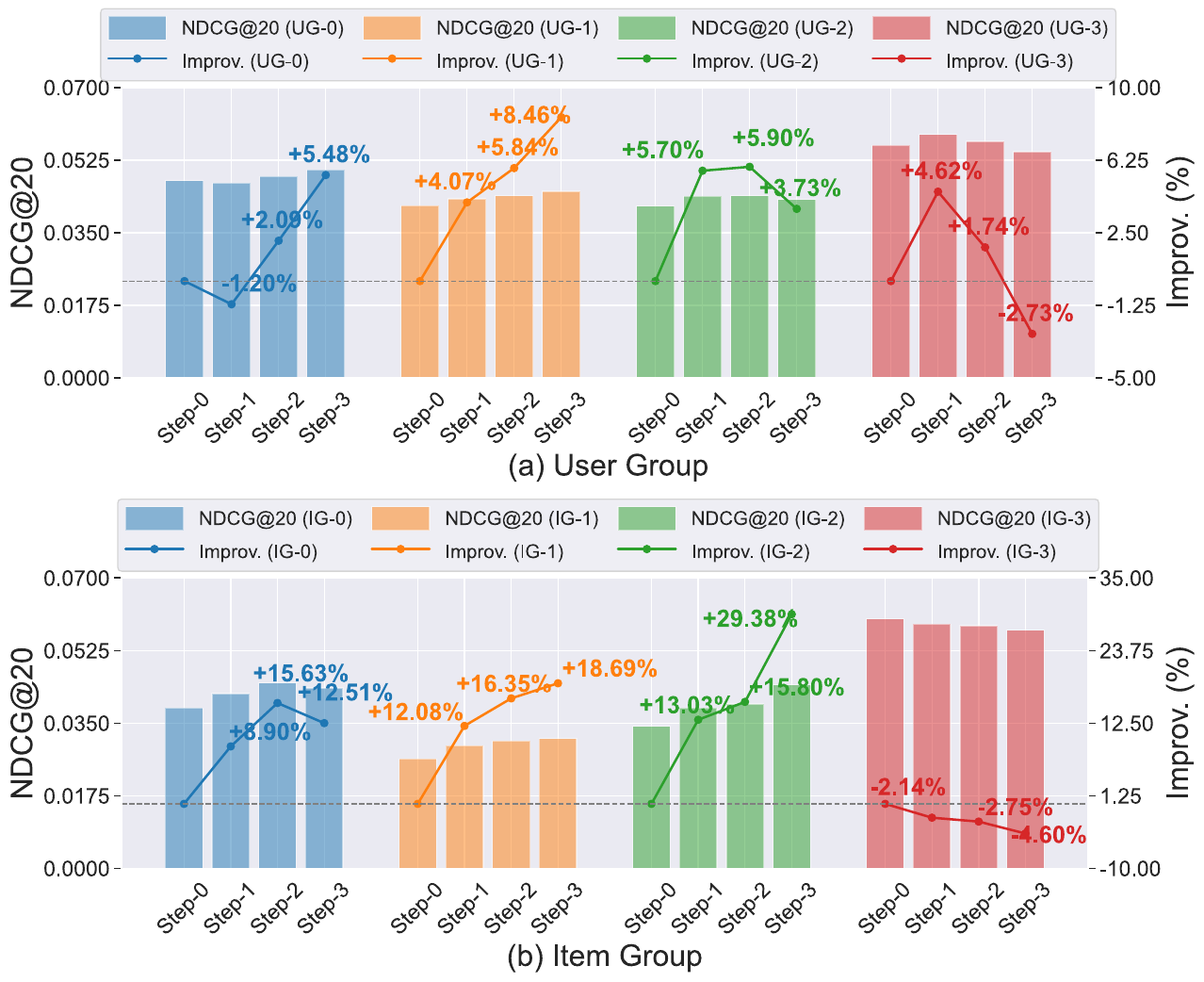}
  \caption{Robustness study w.r.t different user and item subgroups on Yelp dataset. `Step-$x$' represents the recommendation performance at the $x$-th reasoning step. `UG' and `IG' denote User and Item Group, respectively, where higher group numbers indicate longer sequences and more popular items.}
  \label{fig:group}
\end{figure}

\subsection{Further Analysis}
In this section, we provide a comprehensive evaluation of the proposed ReaRec framework. 
We first conduct an in-depth analysis of how reasoning depth impacts performance across different user and item groups(Sec.~\ref{sec:robustness_analysis}). We then explore the impact of reasoning steps on recommendation performance (Sec.~\ref{sec:performance_variation}) and inference latency (Sec.~\ref{sec:reasoning_time}). Next, we perform a detailed ablation study (Sec.~\ref{sec:ablation_study}) and hyperparameter sensitivity analysis (Sec.~\ref{sec:sensitivity_analysis}). Finally, we investigate the visualization of reasoning hidden states to gain insights into the model's reasoning process (Sec.~\ref{sec:visualization_analysis}).
Unless otherwise specified, we primarily conduct detailed experiments on the PRL method based on SASRec backbone using the Yelp and Video \& Games datasets.

\subsubsection{\textbf{Robustness Analysis Across User and Item Subgroups}}\label{sec:robustness_analysis}
To further analyze the robustness of our proposed ReaRec framework, we split users and items into different subgroups to gain deeper insights into the performance of the multi-step reasoning framework.
Specifically, for users, we divide users into four equal-sized groups based on sequence length: \{UG-0, UG-1, UG-2, UG-3\}, where higher group numbers indicate longer sequences. 
For items, following previous work~\cite{yang2023debiased,tang2024towards}, we group them into four groups based on interaction frequency: \{IG-0, IG-1, IG-2, IG-3\}, where higher group numbers indicate more popular items. We ensure each item group contains the same sample numbers. We fix the reasoning steps for PRL method during training at three, and analyze how recommendation performance changes for different user and item groups as reasoning steps increase during the inference phase.
The detailed experimental results are shown in Fig.~\ref{fig:group}.

We can clearly observe distinct performance trends across different user and item subgroups. 
For short-sequence user groups and unpopular item groups, recommendation quality (NDCG@20) tends to steadily improve as the reasoning steps increase. For example, in the item group IG-1, more reasoning steps bring better performance gains of 12.08\%, 16.35\%, and 18.69\%, respectively.
In contrast, performance tends to decline for users with long interaction sequences and popular items as the reasoning steps increase. We speculate that this is primarily because longer user sequences provide richer contextual information, making it easier to mine interest evolution patterns. Beyond a certain point, additional inference computation fails to yield further performance improvements and even leads to performance degradation due to overthinking.
Similarly, for high-popularity items, their well-trained representations allow the recommender to easily capture collaborative signals, making deeper feature crossing depth less beneficial.
Overall, long-tail users and items usually require more thinking space to reason sparse interaction signals, whereas highly active users and items may not need redundant computational expansion. In the future, it may be necessary to develop differentiated fast and slow reasoning mechanism for different user sequences to further improve overall recommendation performance.

\subsubsection{\textbf{Impact of Reasoning Steps on Recommendation Performance}}\label{sec:performance_variation}
We investigate the variation trend of recommendation performance under different inference steps, that is, we train and perform inference using specified numbers of reasoning steps.
We adopt NDCG@20 as the main evaluation metric. 
We compare the following approaches: 
\textbf{(1) Base}: The original SASRec sequential recommender serves as the baseline without reasoning enhancement; 
\textbf{(2) Naive}: Based on the Base method, we extend it to a multi-step reasoning paradigm, where the last hidden state is autoregressively fed back into the model, and only the final position is used directly as the user representation;
\textbf{(3) RPE:} Building on the Naive approach, we further integrate \textit{Reasoning Positional Embeddings} to bridge the task gap between sequence encoding mode and reasoning mode.
Additionally, we also explore the performance of \textbf{(4) Ensemble Reasoning Learning (ERL)} and \textbf{(5) Progressive Reasoning Learning (PRL)} under multi-step reasoning.

{
\renewcommand{\arraystretch}{1.15}
\begin{table}[t]
  \centering
    \caption{Inference time statistics for different steps. ``Cost Inc.'' is short for Cost Increase, where higher values indicate greater time overhead. Efficiency experiments are conducted on a single A100-40G GPU. Note that the optimal performance typically corresponds to Step-2.}
    \label{tab:time_exp}
  \begin{threeparttable}
    \resizebox{\linewidth}{!}{
      \begin{tabular}{*{7}{c}}
    \toprule
    & Base & Step-1 & Step-2 & Step-3 & Step-4 & Step-5 \\
    \midrule
    SASRec & 5.6761 & 5.7985 & 5.8752 & 5.9305 & 6.0310 & 6.2786 \\
    Cost Inc. & - & 2.16\% & 3.51\% & 4.48\% & 6.25\% & 10.61\% \\
    \hline
    BERT4Rec & 5.6535 & 5.7685 & 5.9174 & 5.9621 & 6.0862 & 6.1224 \\
    Cost Inc. & - & 2.03\% & 4.67\% & 5.46\% & 7.65\% & 8.29\% \\
    \hline
    UniSRec & 5.6061 & 5.6312 & 5.7596 & 5.8732 & 6.0303 & 6.0502 \\
    Cost Inc. & - & 0.45\% & 2.74\% & 4.76\% & 7.57\% & 7.92\% \\
    \hline
    MoRec & 5.6638 & 5.7143 & 5.8391 & 5.9565 & 5.9659 & 5.9812 \\
    Cost Inc. & - & 0.89\% & 3.10\% & 5.17\% & 5.33\% & 5.60\% \\
    \bottomrule
\end{tabular}
    }
    \begin{tablenotes}
      \item Note: All time units are in second (s).
  \end{tablenotes}
\end{threeparttable}
\vspace{-0.5em}
\end{table}
}

\begin{figure}
  \centering
  \includegraphics[width=\linewidth]{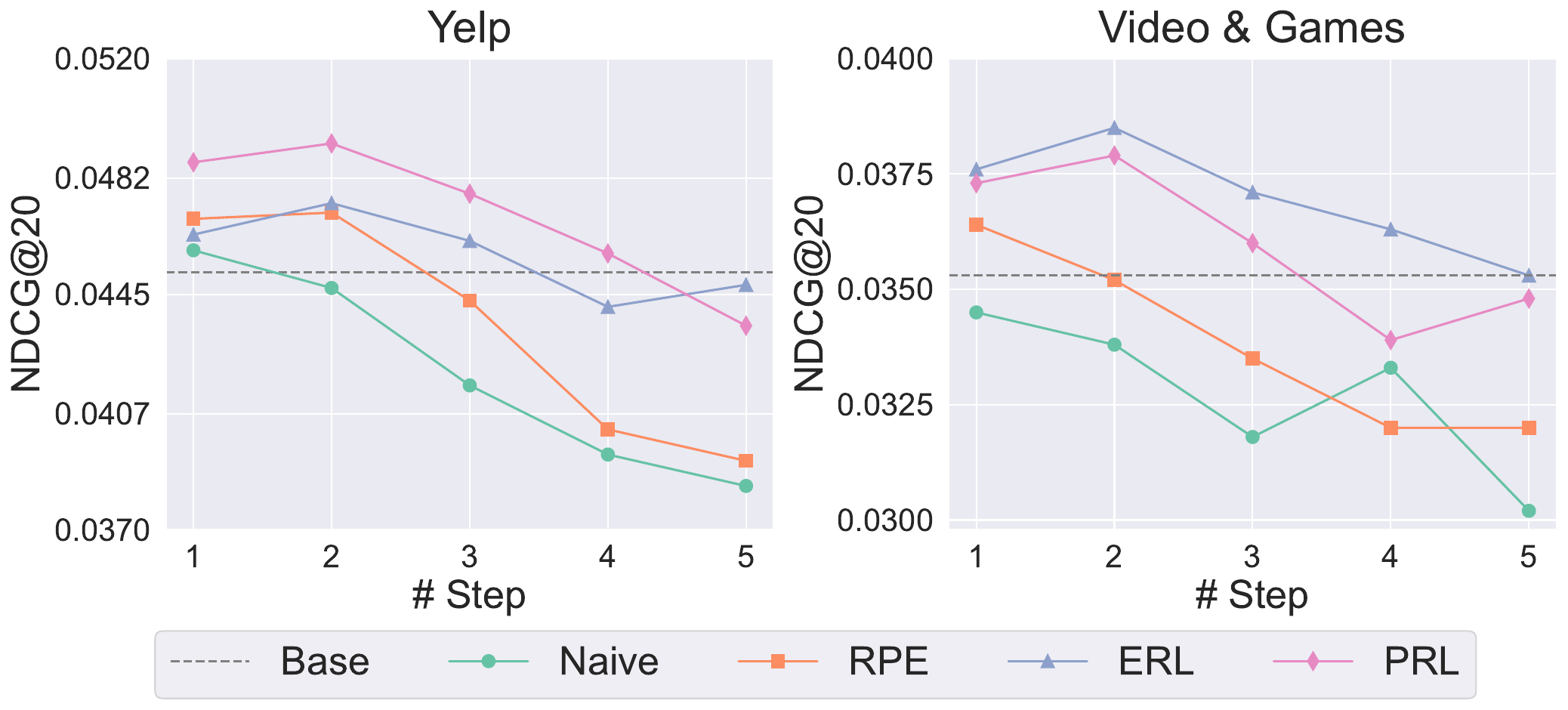}
  \caption{The performance variation trend of different methods under different reasoning steps.}
  \label{fig:step_trend}
\end{figure}

\begin{figure*}[t]
  \captionsetup{aboveskip=0pt}
  \centering
  \includegraphics[width=\linewidth]{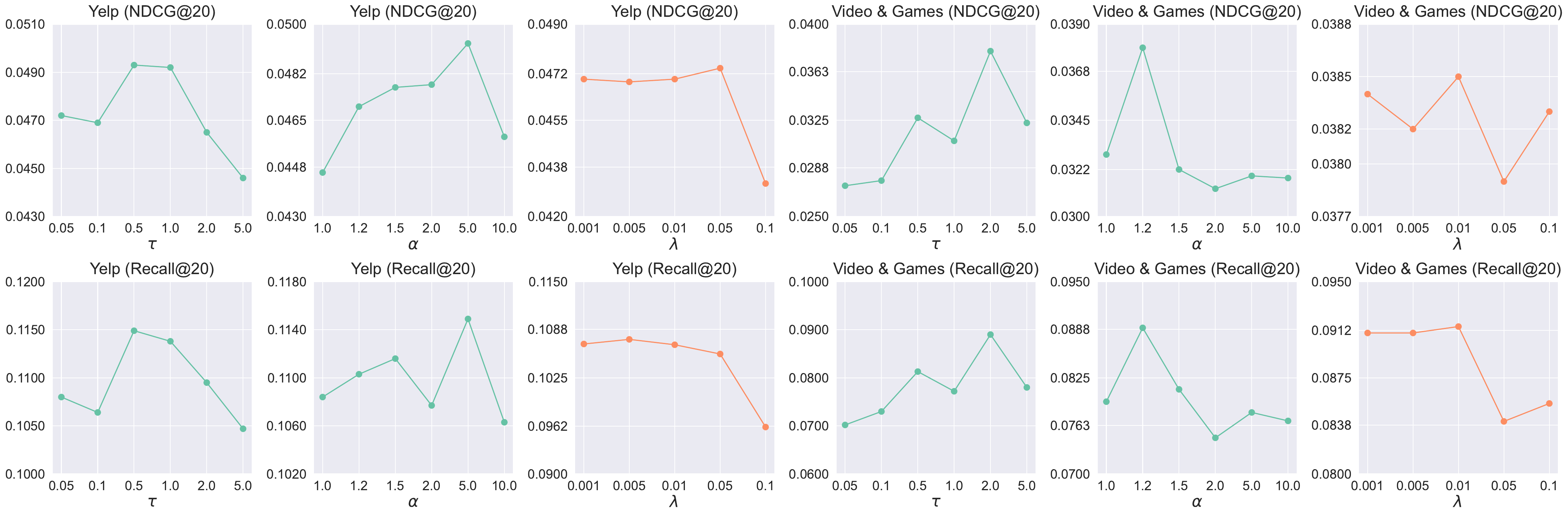}
  \caption{Performance comparison w.r.t. different hyperparameters, including base temperature $\tau$, temperature decay rate $\alpha$, and KL regularization strength $\lambda$. The \textcolor[HTML]{7DC0A7}{green} and \textcolor[HTML]{ED936B}{orange} lines represent the PRL and ERL methods, respectively.}
  \label{fig:param_study}
\end{figure*}

As shown in Fig.~\ref{fig:step_trend}, the Naive method, which lacks a specialized design, does not yield performance improvements and even underperforms compared to the base model. This is likely due to the model's inability to distinguish between sequence encoding and the reasoning phases.
Introducing reasoning positional embeddings (+RPE) effectively mitigates this task gap, yielding obvious performance gains.
However, simply optimizing cross-entropy loss on the final-step output does not provide adequate supervision guidance for the intermediate reasoning states, potentially leading to reasoning pattern degradation and error accumulation.
In contrast, our ERL and PRL methods significantly alleviate these issues by explicitly injecting stepwise supervision signals, reducing the optimization difficulty to some extent.
Notably, as the number of inference steps increases, we observe a consistent performance decline across all methods. This suggests that excessive reasoning may trigger ``\textbf{overthinking}''---simple user interaction patterns may not require intensive latent reasoning. 
Moreover, considering the post-hoc optimal step analysis in Fig.~\ref{fig:upper_bound}, developing an adaptive inference depth selection mechanism to balance reasoning depth and user sequence complexity presents a highly meaningful direction for future research.

\begin{figure}
  \captionsetup{belowskip=-10pt}
  \centering
  \includegraphics[width=\linewidth]{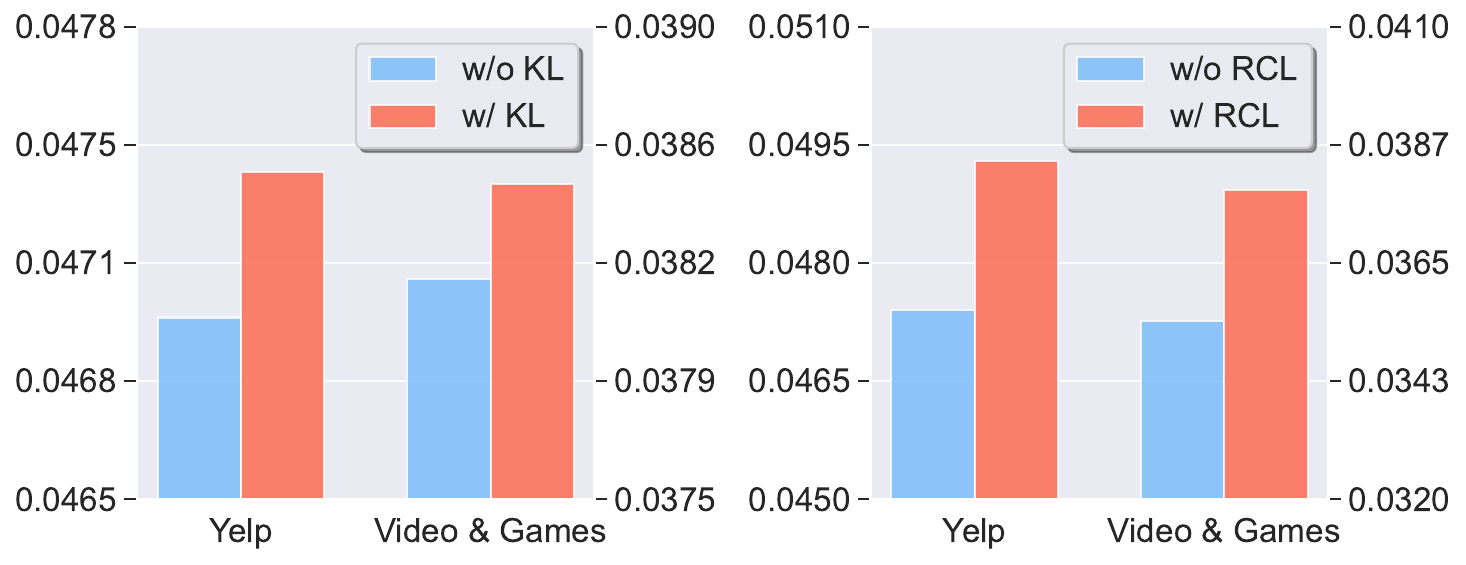}
  \caption{Ablation study for key components in ERL and PRL.}
  \label{fig:ablation_study}
\end{figure}

\subsubsection{\textbf{Impact of Reasoning Steps on Inference Latency}}\label{sec:reasoning_time}
Our ReaRec framework's expanded computational demands during inference introduce additional overhead. 
To evaluate this, we use the PRL method as an example, measuring the time cost on the test set as reasoning steps increase, as shown in Table~\ref{tab:time_exp}. The results indicate that, despite adopting a recurrent autoregressive inference mechanism, the extra latency remains manageable. This efficiency stems from KV Caching technique, which significantly reduces attention computation complexity from $O(N^2)$ to $O(N)$ by reusing key and value vectors of past steps, thereby effectively minimizing redundant calculations.
Further analysis with Fig.~\ref{fig:step_trend} reveals that our approaches generally achieve optimal performance at two reasoning steps. This means that our method increases performance by an average of 7.49\% across all metrics with only a modest latency overhead of 3.51\%, which is acceptable and practical for real-world deployment in industrial recommender systems. These results suggest that our efficient ReaRec framework holds great promise for real-world applications.

\subsubsection{\textbf{Ablation Study}}\label{sec:ablation_study}

In this section, we present the ablation study of our proposed method. Specifically, we focus on two key components: \textbf{(1) KL regularization term (KL)} in the ERL approach (Sec.~\ref{sec:kl}) and \textbf{(2) Reasoning-aware Contrastive Learning (RCL)} in the PRL method (Sec.~\ref{sec:rcl}).
Specifically, we conduct ablation studies by removing the auxiliary loss terms from both methods and evaluate their performance on NDCG@20. 

As shown in Fig.~\ref{fig:ablation_study}, the experimental results clearly indicate that the ERL method without KL regularization performs worse than the full model, suggesting that the model probably suffers from pattern degradation in reasoning states, leading to highly homogeneous outputs. 
Similarly, the PRL method without RCL also yields suboptimal recommendation performance. While progressive temperature scheduling helps adjust the learned distribution sharpness across different steps, the absence of robust inference mechanisms prevents the recommender from self-correcting deviations in intermediate reasoning states. As a result, it struggles to effectively approximate the user's true preference distribution.

\subsubsection{\textbf{Sensitivity Analysis}}\label{sec:sensitivity_analysis}
In this section, we examine the effects of three key hyperparameters, $\tau$, $\alpha$, and $\lambda$ on the Yelp and Video \& Games datasets. Here, $\tau$ and $\alpha$ represent the base temperature and progressive temperature decay rate in the PRL method, respectively, while $\lambda$ denotes the KL regularization strength in the ERL method. We next analyze how variations in each hyperparameter influence model performance, with the experimental results shown in Fig.~\ref{fig:param_study}.

\begin{figure*}[t]
  \centering
  \begin{subfigure}{0.245\textwidth}
    \centering
    \includegraphics[width=\linewidth]{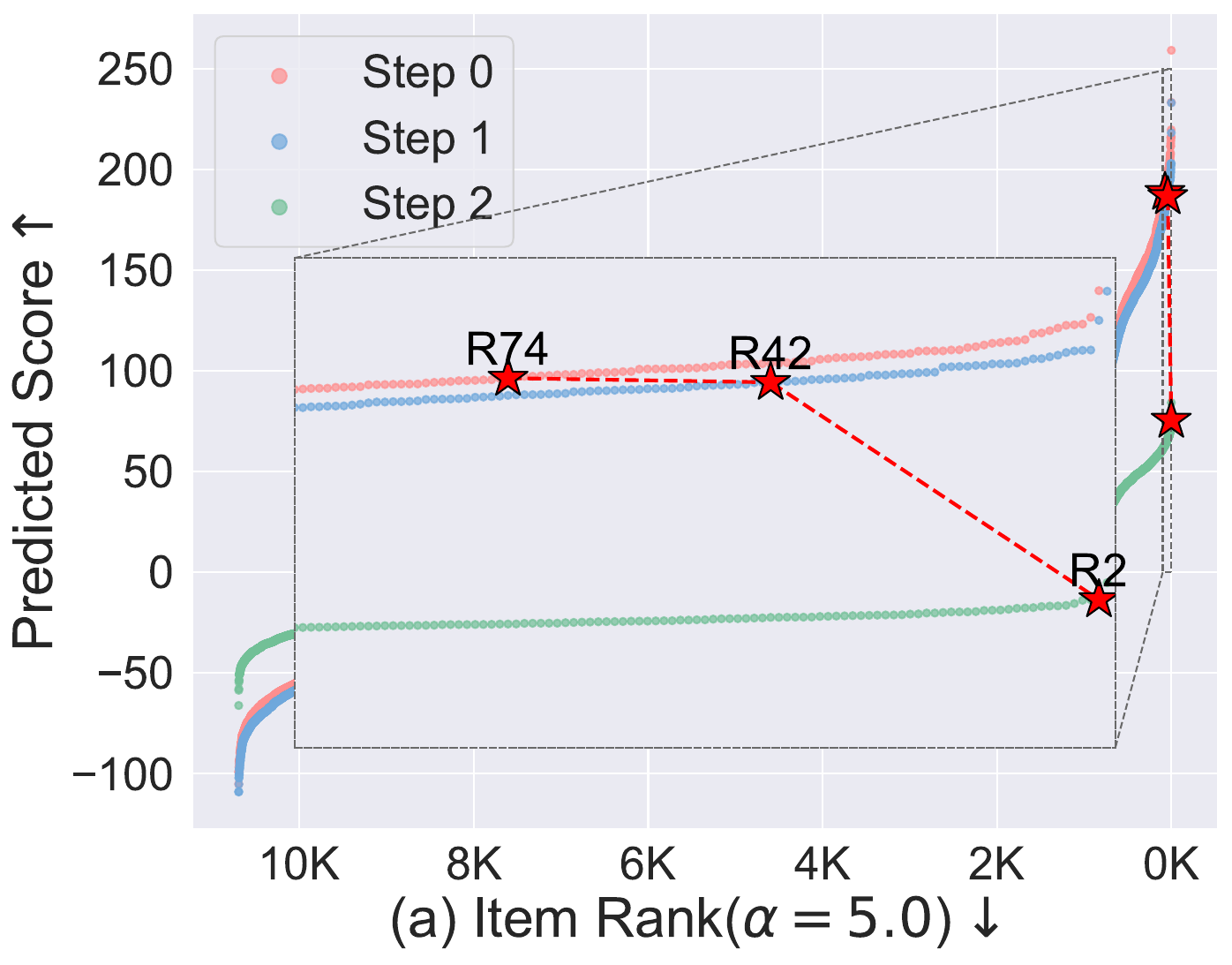}
  \end{subfigure}
  \hfill
  \begin{subfigure}{0.245\textwidth}
    \centering
    \includegraphics[width=\linewidth]{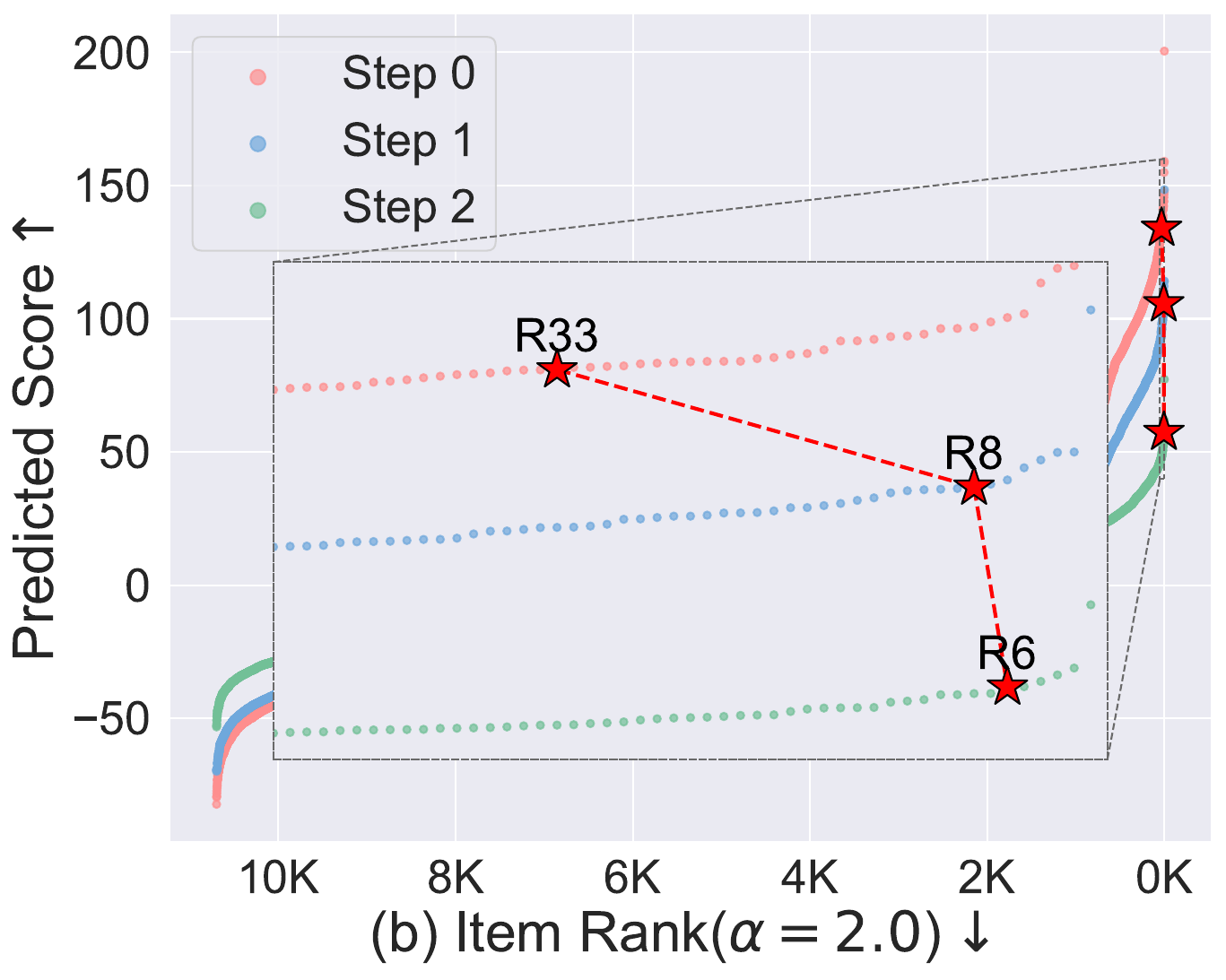}
  \end{subfigure}
  \hfill
  \begin{subfigure}{0.245\textwidth}
    \centering
    \includegraphics[width=\linewidth]{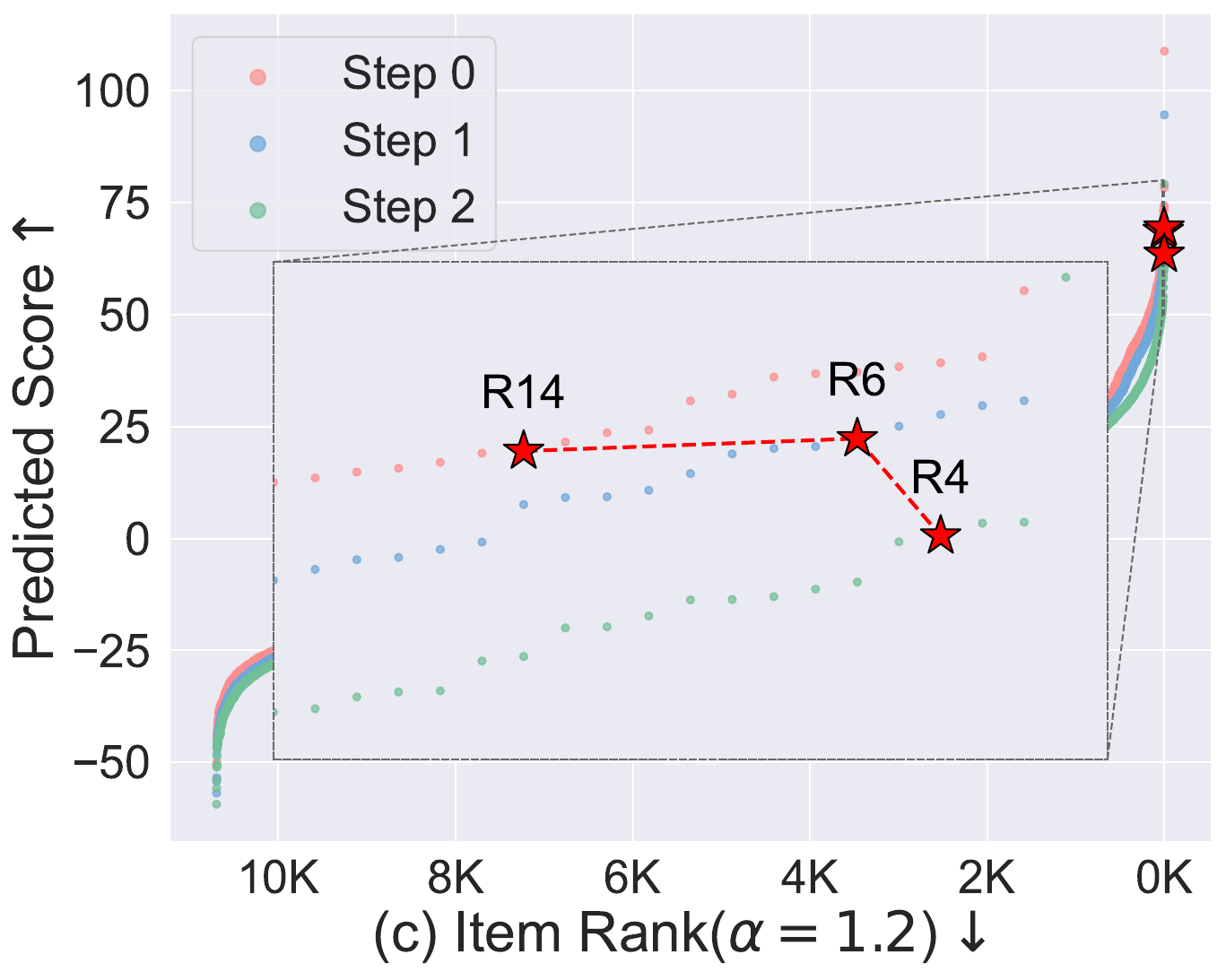}
  \end{subfigure}
  \hfill
  \begin{subfigure}{0.245\textwidth}
    \centering
    \includegraphics[width=\linewidth]{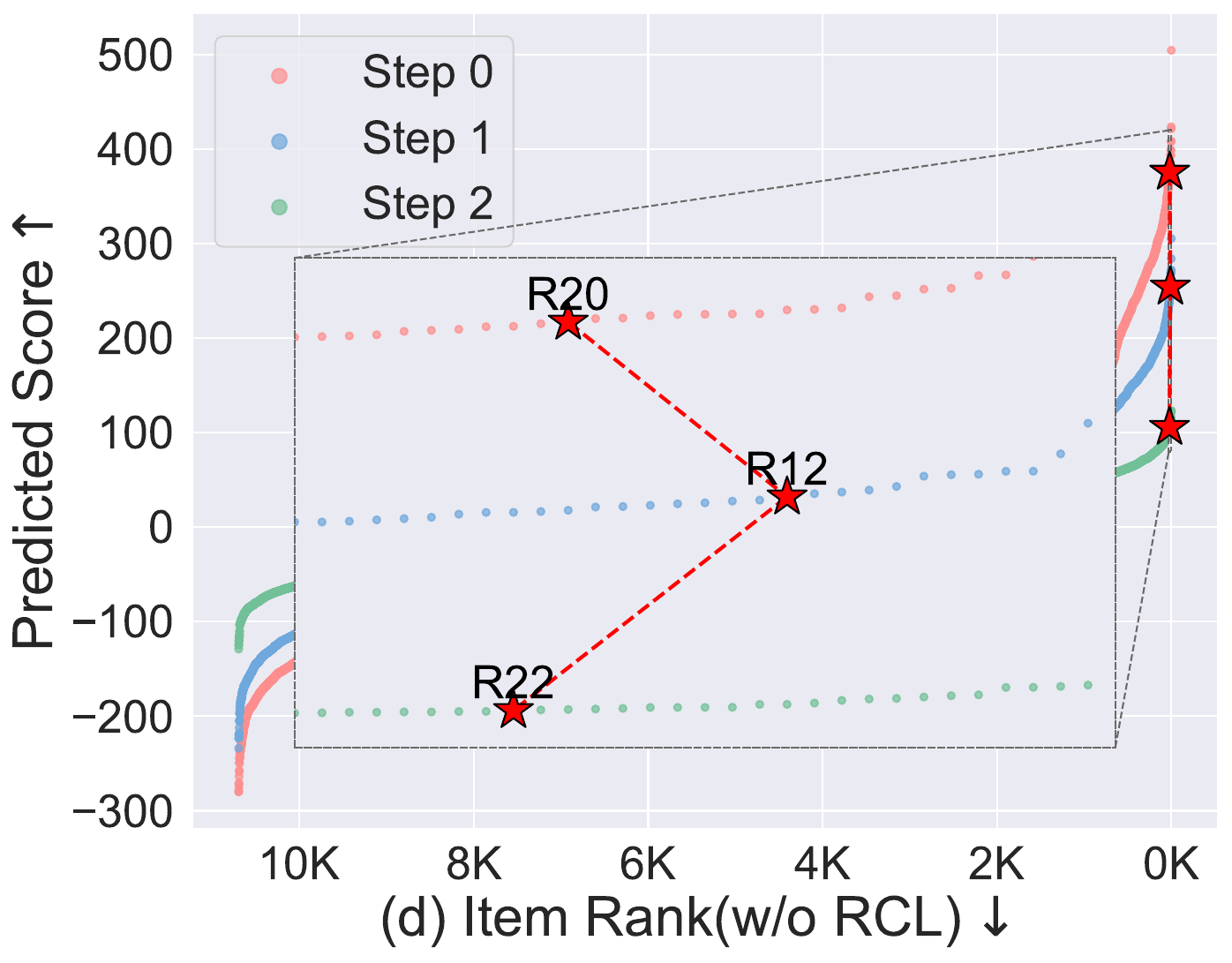}
  \end{subfigure}
  \caption{Case study on rank changes of target item across different reasoning steps. `Rx' represents the predicted rank of the target item, \emph{e.g.}, `R42' indicates the predicted score of the target item ranks 42nd among all candidate items.}
  \label{fig:item_rank}
\end{figure*}

\textbf{\textit{Performance Comparison w.r.t Base Temperature $\bm{\tau}$ in PRL.}}
By tuning the base temperature $\tau$ across $\{0.05, 0.1, 0.5, 1.0, 2.0, 5.0\}$, we can observe that as $\tau$ increases, model performance gradually improves. This suggests that overly sharp probability distribution does not align with users' potential preference distributions. In other words, forcing the model to learn extreme positive and negative sample preferences from noisy interaction data hinders generalization ability.
However, too large base temperatures also lead to degraded recommendation performance. We hypothesize that a large $\tau$ value may blur the ranking differences among candidate items, making it harder for the recommender to learn meaningful sequential patterns. Thus, setting a moderate $\tau$ is crucial for achieving satisfactory performance.

\begin{figure}[t]
  \centering
  \begin{subfigure}[b]{0.49\linewidth}
    \centering
    \includegraphics[width=\linewidth]{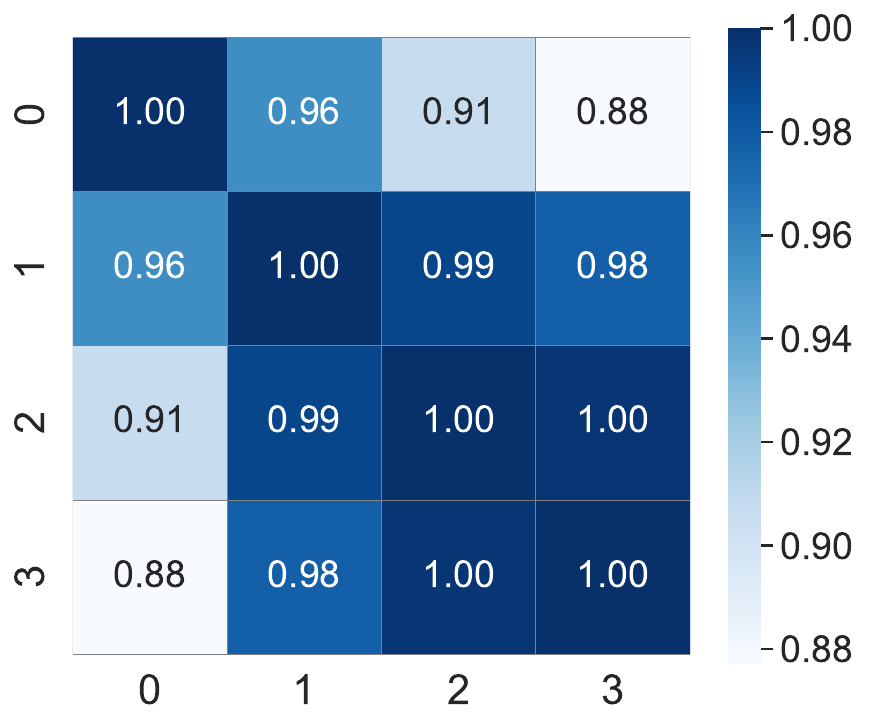}
    \caption{RPE}
    \label{fig:prl_similarity}
  \end{subfigure}
  \hfill
  \begin{subfigure}[b]{0.48\linewidth}
    \centering
    \includegraphics[width=\linewidth]{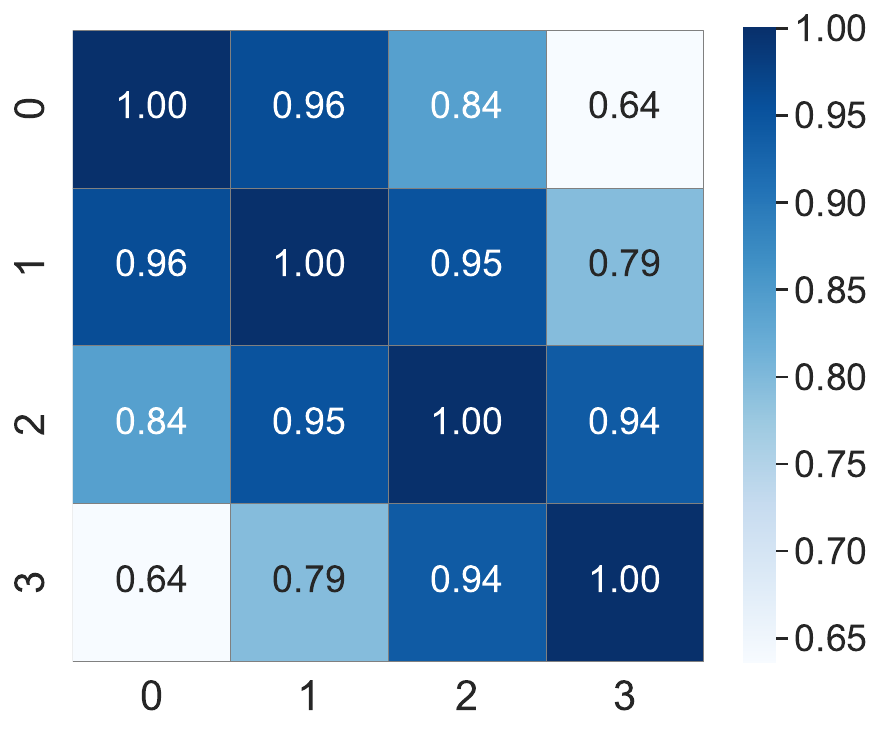}
    \caption{PRL}
    \label{fig:pos_similarity}
  \end{subfigure}
  \begin{subfigure}[b]{0.49\linewidth}
    \centering
    \includegraphics[width=\linewidth]{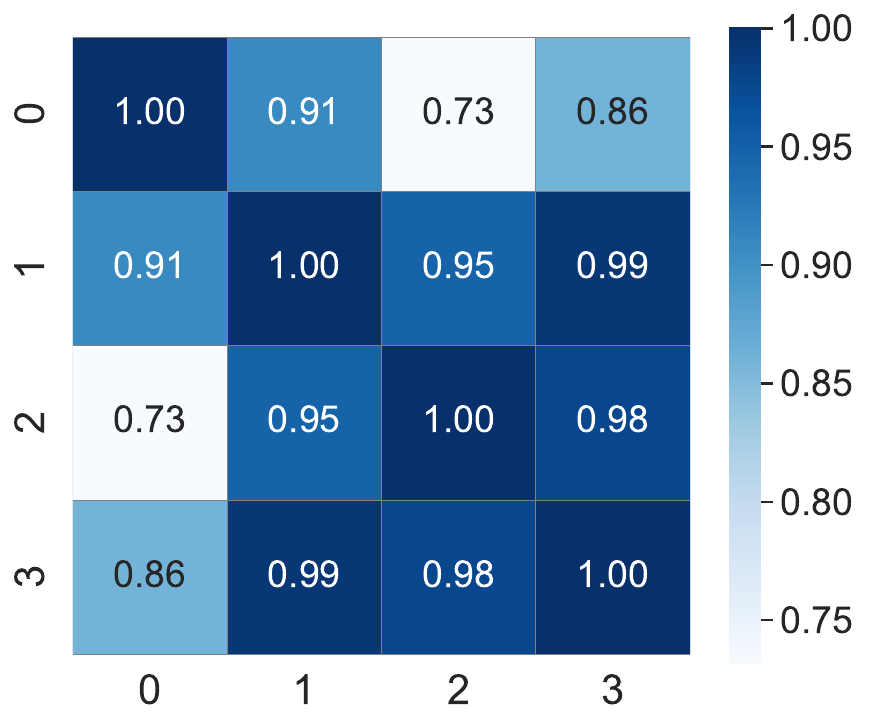}
    \caption{ERL w/o KL}
    \label{fig:prl_similarity2}
  \end{subfigure}
  \hfill
  \begin{subfigure}[b]{0.49\linewidth}
    \centering
    \includegraphics[width=\linewidth]{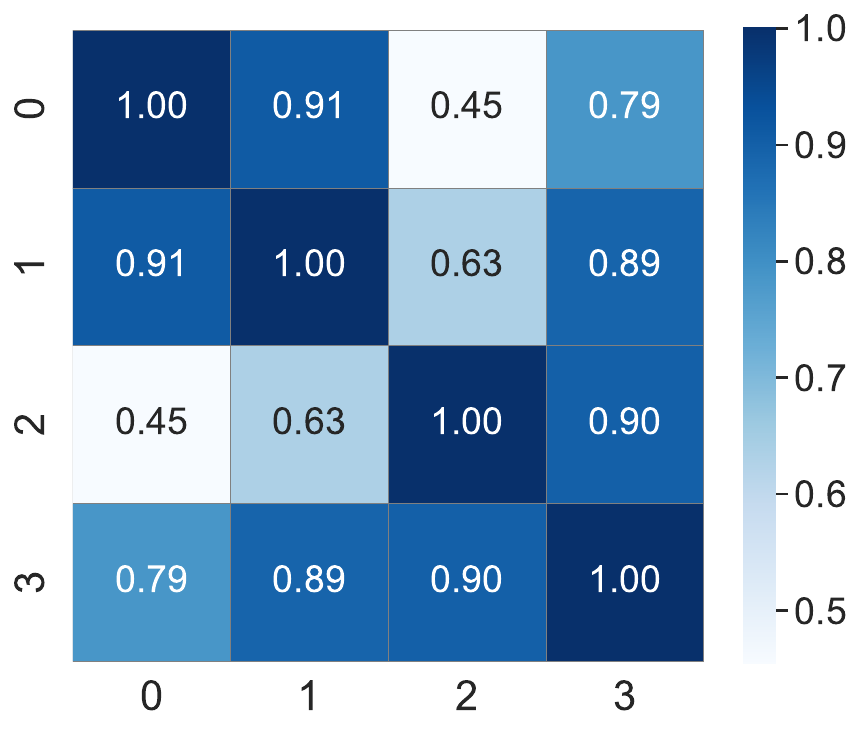}
    \caption{ERL}
    \label{fig:pos_similarity2}
  \end{subfigure}
  \caption{Visualization of similarity in multi-step reasoning hidden states for different methods.}
  \label{fig:heatmaps}
  \vspace{-0.5cm}
\end{figure}

\textbf{\textit{Performance Comparison w.r.t Temperature Decay Rate $\bm{\alpha}$ in PRL.}}
To enable the model to learn a more precise user preference distribution, we introduce a progressive temperature annealing mechanism controlled by temperature decay rate $\alpha$, adjusting the sharpness of the learned distribution at different reasoning steps. 
We vary it within $\{1.0, 1.2, 1.5, 2.0, 5.0, 10.0\}$ to observe the performance changes. 
A consistent finding is that a moderate $\alpha$ usually achieves the best performance, while too small and too large decay rates lead to suboptimal results. 
This is as expected, as $\alpha$ is too small (in the extreme case, $\alpha=1.0$), the score distributions learned at different reasoning steps remain the same, causing the model to take shortcuts like replicating the prior reasoning state. Such pattern collapse prevents the model from leveraging reasoning enhancement in inference.
On the other hand, overly high $\alpha$ (e.g., $\alpha=10.0$) still leads to performance degradation. This is because, under our exponential temperature decay strategy, an aggressive temperature change triggers a rapid distribution sharpness transition from smooth to sharp distribution, disrupting the model's curriculum-style reasoning process.
Therefore, choosing an approximate temperature decay rate is critical for reducing the model's optimization difficulty.

\begin{figure}
  \centering
  \begin{subfigure}[b]{\linewidth}
    \includegraphics[width=\linewidth]{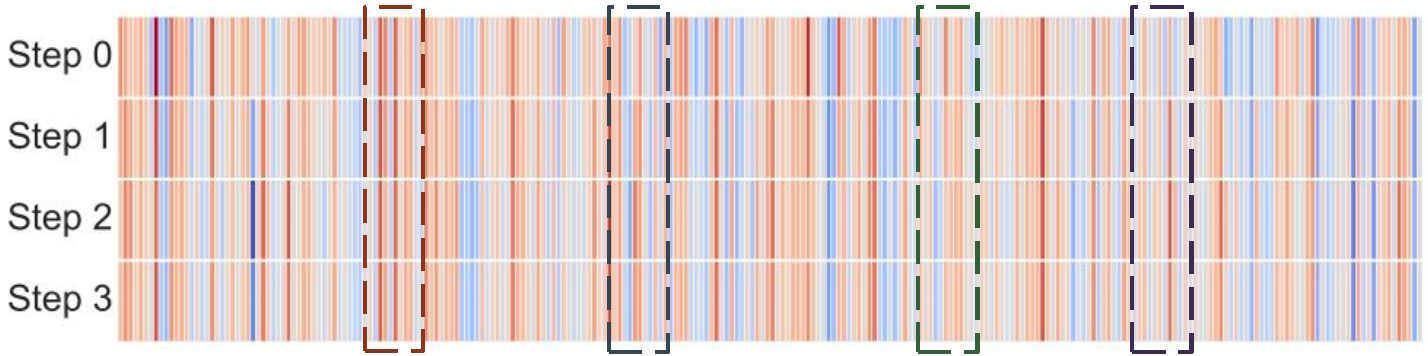}
    \caption{ERL w/o KL}
  \end{subfigure}
  \begin{subfigure}[b]{\linewidth}
    \includegraphics[width=\linewidth]{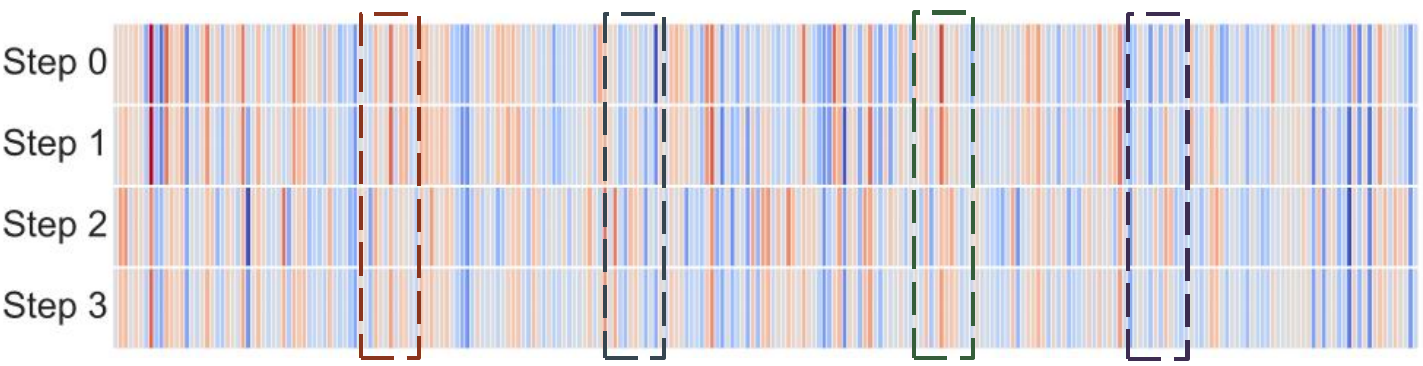}
    \caption{ERL}
  \end{subfigure}
  \vspace{-0.5cm}
    \caption{The embedding visualization of the full ERL method \emph{vs.} its ablated version without KL regularization. Dashed boxes highlight high similarity between different reasoning steps (Step 0 $\sim$ Step 3) in the ablated version.}
  \label{fig:embedding}
\end{figure}

\textbf{\textit{Performance Comparison w.r.t KL Regularization Strength $\bm{\lambda}$ in ERL.}}
In our ensemble reasoning learning method, we utilize the KL regularization coefficient $\lambda$ to balance reasoning diversity.
Specifically, we explore the impact of different regularization strengths by varying $\lambda$ within the range $\{0.001, 0.005, 0.01, 0.05, 0.1\}$.
From Fig.~\ref{fig:param_study}, we can observe that the model is usually not sensitive to the $\lambda$. 
However, the recommendation performance drops significantly when $\lambda$ exceeds a certain threshold (\emph{e.g.}, 0.05). We attribute this to that enforcing the model to learn excessively divergent sequential patterns across multi-step reasoning might actually disrupt the sequential modeling capability. 
Although our designed KL regularization aims to encourage the model to explore diverse reasoning paths, too strong regularization may dominate gradient optimization, increasing the optimization challenges and ultimately leading to performance degradation.

\subsubsection{\textbf{Embedding Visualization Analysis}}\label{sec:visualization_analysis}

To analyze the hidden state dynamics during reasoning, we visualize the similarity heatmaps of multi-step reasoning outputs for different methods, as shown in Fig.~\ref{fig:heatmaps}.
Specifically, by comparing Fig.~\ref{fig:heatmaps}(a) and Fig.~\ref{fig:heatmaps}(b), it is obvious that the RPE variant exhibits high homogeneity in reasoning states. For instance, the similarity scores between the final output and the previous two steps are almost identical \emph{i.e.}, 1.00 and 0.98, which confirms the reasoning pattern degradation issue claimed before. 
In contrast, by incorporating a progressive reasoning learning approach, PRL effectively leverages reasoning-enhanced computation for performance improvement. 
The ERL method demonstrates analogous issues, where KL regularization encourages the model to capture diverse sequential patterns through aggregating multi-order feature crossing. 
Additionally, we visualize the specific reasoning representations in Fig.~\ref{fig:embedding}, where we can observe that the ERL method without KL constraint reveals more overlapping patterns across different reasoning steps. This further validates that our proposed methods can effectively address core challenges in multi-step reasoning sequential models.

\begin{figure}[t]
  \captionsetup{belowskip=-8pt}
  \centering
  \includegraphics[width=\linewidth]{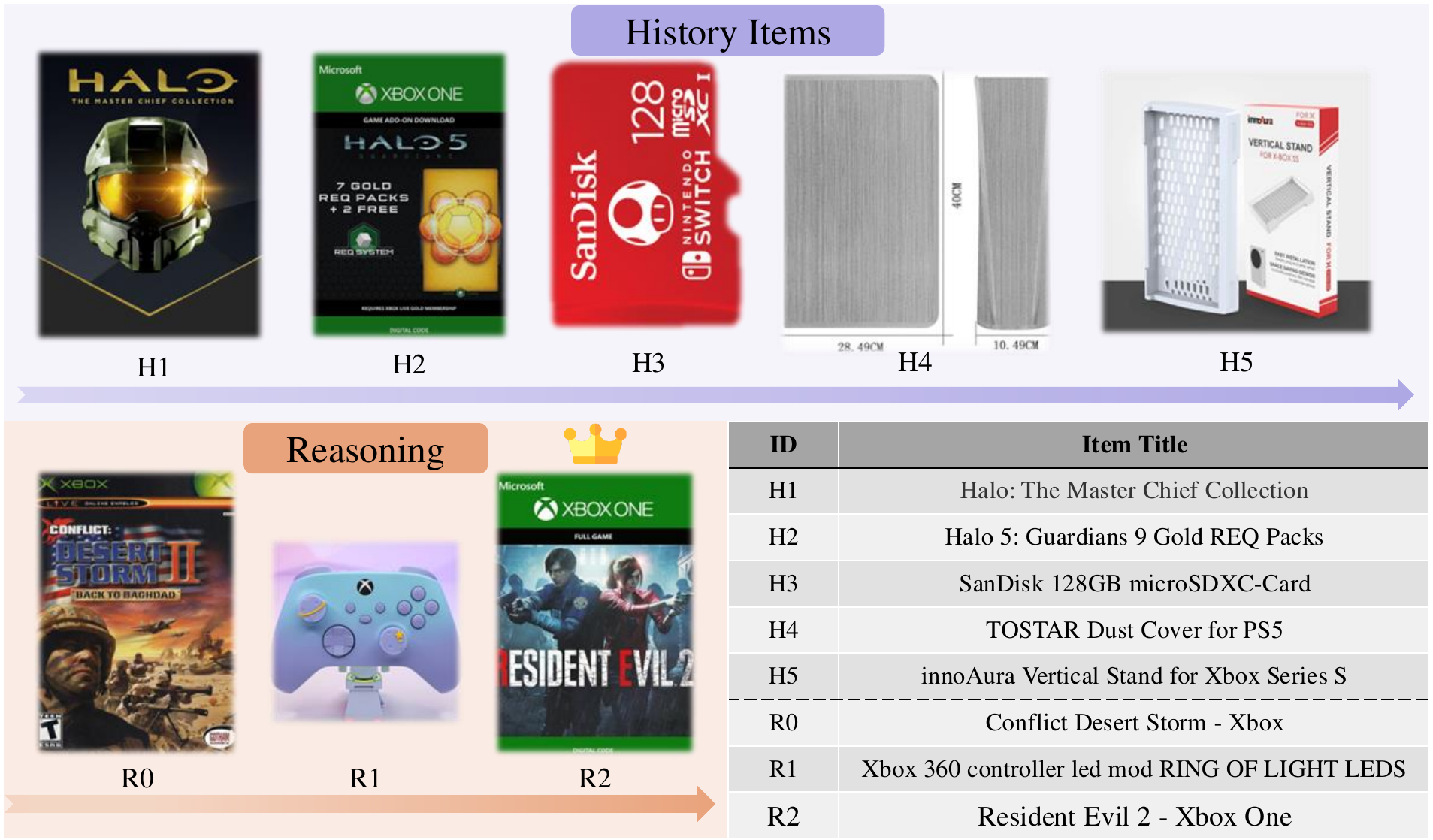}
  \caption{Case study of multi-step inference on the Video \& Games Dataset. `H$x$' represents historical items, with smaller $x$ indicating more recent interactions. `R$x$' represents the top-1 recommended items at the $x$-th reasoning step, with larger $x$ indicating later reasoning steps.}
  \label{fig:case_study}
\end{figure}

\subsection{Case Studies}
In this section, to better demonstrate the benefits of our reasoning-enhanced sequential recommendation, we present illustrative cases showing the rank change trends of target items during mulit-step inference, along with a specific example from Video \& Games dataset.

\subsubsection{\textbf{Rank Change Analysis of Target Items}}
We evaluate target item ranking trajectories on the Yelp dataset using PRL methods with varying temperature decay coefficient ($\alpha$) and an ablated version without RCL.
As shown in Fig.~\ref{fig:item_rank}, we observe that the full PRL method progressively improves the target item ranking within the overall candidate pool as the depth of reasoning increases, aligning with our expectations.
Additionally, we find that for smaller $\alpha$, the score distribution across different inference steps transitions smoothly, whereas larger $\alpha$ induces distribution changes more aggressively, which is consistent with our analysis in Sec.~\ref{sec:sensitivity_analysis}.
Moreover, the ablated version without RCL leads to reasoning errors where increasing the number of reasoning steps incorrectly pushes the target item further down the ranking (\emph{e.g.}, target item rank drops from \#12 at step 1 to \#22 at step 2 in Fig.~\ref{fig:item_rank}(d)).

\subsubsection{\textbf{Case Study in Real-world Recommendation Scenario}}
We present a case study to illustrate the stepwise preference refinement effect of the PRL method, as shown in Fig.~\ref{fig:case_study}.
To be specific, the user previously purchased \textit{Halo} and \textit{Halo 5}, two \textit{First-Person Shooter (FPS)} games for the XBox-One platform on Amazon. After that, the user bought related accessories, \emph{i.e.}, a memory card, a dust cover, and a stand. Next, the corresponding top-1 recommended items are given by the multi-step reasoning outputs, denoted as R0, R1, and R2, respectively.
At step R0, the model successfully captures the user's preference for FPS games on the XBox platform. However, this recommendation (\emph{i.e.}, \textit{Conflict Desert Storm}) lacks timeliness and may not align with a gaming enthusiast's tendency to prefer newer releases.
At step R1, the model adjusts by recommending a game controller, reflecting the user's recent purchase habits (\emph{i.e.}, gaming accessories). However, this recommendation remains suboptimal, as it only reflects collaborative relevance rather than sequential characteristics (typically, users phase controllers before accessories like stands) and lacks recommendation diversity (as recent purchases were all accessories).
Surprisingly, at the final inference step, the model recommends \textit{Resident Evil 2}, a newly released shooter game that matches the actual target item and aligns well with the true preference, further validating how recurrent reasoning resolves ambiguity by integrating temporal context, collaborative relevance, and output diversity.

\section{Related Work}
\textbf{Sequential Recommendation.}
Sequential recommendation, as one of the core tasks within the field of \textit{Recommender Systems (RS)}, aims to predict the next item with which a user may interact by modeling behavior patterns and evolving user interests from user-item interaction sequences~\cite{boka2024survey,wang2019sequential,fang2020deep}.
Early studies focused on item-to-item collaborative transitions, typically employing Markov Chain-based matrix factorization methods~\cite{he2016fusing,rendle2010factorizing} to achieve next-item predictions.
With the advent of deep learning, researchers began exploring various sequential modeling architectures, such as Recurrent Neural Networks (RNNs)~\cite{hidasi2015session,hidasi2016parallel,quadrana2017personalizing}, Convolutional Neural Networks (CNNs)~\cite{tang2018personalized,yuan2019simple,yan2019cosrec}, and Transformer~\cite{kang2018self,chen2019behavior,yang2022multi}, to enhance modeling capabilities.
Specifically, GRU4Rec~\cite{hidasi2015session} first introduced the GRU networks for session-based recommendation. Caser~\cite{tang2018personalized} applied convolutional operations by treating the item sequence embedding matrix as an ``image'' to extract multi-level interaction features. With the emergence of the Transformer architecture, numerous studies in sequential recommendation shifted toward self-attention-based models. For instance, SASRec~\cite{kang2018self}, a classic sequential recommendation baseline, incorporated self-attention to automatically learn the importance weights of historical items. BERT4Rec~\cite{sun2019bert4rec} further advanced this by adopting bidirectional encoding to capture more contextual dependencies from both directions of the sequence.
Furthermore, to address data sparsity and cold-start issues, another line of research focused on leveraging common item attributes (such as text and images) to learn universal item and sequence representations~\cite{hou2022towards,yuan2023go,ding2021zero,hou2023learning}. For example, UniSRec~\cite{hou2022towards} utilized multi-domain recommendation data to learn transferable sequential models.
However, existing methods remain constrained by reasoning-free forward computation. In this paper, we further explore the potential reasoning capabilities of sequential recommenders by extending the multi-step computational depth at inference time.

\textbf{Inference-time Reasoning.}
As the scaling law of large language models (LLMs) at the training stage has gradually reached its bottleneck~\cite{chen2025towards,snell2024scaling,kumar2025llm}, researchers have shifted their focus toward inference-time scaling.
Notably, OpenAI's O1 series~\cite{jaech2024openai}, Qwen's QwQ series~\cite{qwq-32b-preview}, and DeepSeek's R1 series~\cite{guo2025deepseek} have emerged as key milestone works, marking the transition from conversational AI to reasoning-intensive AI and opening promising pathways toward achieving Artificial General Intelligence (AGI).
These works leverage emerging long Chain-of-Thought (CoT) mechanisms to reveal excellent test-time scaling phenomena---where increased computational power (via generating more tokens)  during inference substantially improves the model's problem-solving abilities~\cite{wei2022chain,besta2024graph,yao2023tree,yang2024buffer,wang2022self,chu2023navigate,zhang2025and}.
Extended reasoning space enables depth-scalable exploration, manifesting self-reflection capabilities (e.g., ``Aha Moments'') that surpass traditional short-chain reasoning limitations~\cite{chen2025towards,ji2025test}.
Compared with prior methods constrained by token-by-token generation in the discrete language spaces, which restricts the model's expressive ability, another research direction focuses on implicit chain of thought reasoning in latent spaces, achieving both efficiency and performance gains in large language models~\cite{hao2024training,xu2025softcot,geiping2025scaling,chen2024language} and multimodal foundation models~\cite{shen2025efficient,he2024multi}.
For example, Coconut~\cite{hao2024training} introduces continuous thinking in the latent reasoning space of LLMs, while Heima~\cite{shen2025efficient} compresses the entire multimodal CoT process into a single high-level thinking token (\emph{i.e.}, <CoT>) to eliminate redundant intermediate token generation.
Inspired by this think-before-action paradigm, we pioneer the exploration of implicit reasoning-enhanced sequential recommendation framework in this paper, proposing two lightweight reasoning learning methods to push the performance ceiling of sequential recommenders.

\section{Conclusion and Future Work}
\subsection{Conclusion} 

In this work, we pioneer the integration of deep reasoning into sequential recommendation by introducing \textbf{ReaRec}, a novel inference-time computing framework inspired by the \textit{think-before-action} paradigm. Unlike traditional direct inference models, ReaRec expands computational depth through multi-step implicit reasoning, enabling the SeqRec model to think before recommendation. We also propose two lightweight learning strategies to address the challenges of multi-step reasoning-process optimization: Ensemble Reasoning Learning (ERL) and Progressive Reasoning Learning (PRL), which enhance reasoning robustness and effectiveness. Extensive experiments across five real-world datasets validate the effectiveness and generalizability of our proposed ReaRec. Notably, ReaRec not only improves performance for long-tail users and items but also raises the performance ceiling of existing SeqRec backbones by up to 50\% with post-hoc optimal step selection, highlighting the untapped potential of ReaRec for sequential recommendation. We believe our work opens a promising direction for future research at the intersection of reasoning and recommendation.

\subsection{Future Work} 
While our proposed simple inference-time computational strategies successfully unlock the reasoning potential of sequential recommenders and achieve promising performance gains, this work serves primarily as an \textbf{initial exploratory effort}. Consequently, we have also identified some immediate challenges and opportunities for future research:
\begin{itemize}[leftmargin=*]
  \item \textbf{\textit{Adaptive Inference Depth Selection.}} As shown in Fig.~\ref{fig:group}, we observe that while our method effectively improves recommendation performance to cold-start users and long-tail items, it paradoxically induces \textbf{performance degradation} for high-activity users and popular items. We attribute this to the \textbf{overthinking phenomenon}---the additional computational steps provide negligible benefits for well-learned patterns, as their preferences may be sufficiently captured through shallow reasoning-free inference. Moreover, complemented by the post-hoc optimal step analysis in Fig.~\ref{fig:upper_bound}, which illustrates the performance upper bound corresponding to the optimal reasoning step, it becomes evident that there is still a significant gap between the model's current performance and the theoretical upper bound. Therefore, how to develop an \textbf{adaptive inference depth selection policy} to balance computational depth and sequence complexity is an open research direction.
  \item \textbf{\textit{Parameter Disentanglement Between Encoding and Reasoning.}} Our current ReaRec framework adopts an implicit reasoning mechanism similar to large reasoning models, where the item sequence encoding phase shares parameters with the reasoning computations. While this design ensures parameter efficiency, it creates task ambiguity---the same neural modules have to simultaneously handle two distinct objectives: \textbf{(1) precise item presentation learning} and \textbf{(2) multi-step forward reasoning}. Although we propose reasoning position embeddings (\emph{cf.} Sec.~\ref{sec:rearec}) to alleviate this issue, the suboptimal performance trajectories (improvement followed by decline as steps increase shown in Fig.~\ref{fig:step_trend}) suggests our solution may not be optimal. A promising future direction is to explore parameter decoupling between item encoding and deep sequential reasoning at the model level. This separation could potentially reduce task interference, allowing for more specialized representation learning and better adaption to multi-step inference, ultimately leading to improved recommendation quality.
  \item \textbf{\textit{The Missing Inference-time Scaling Law.}} In the field of large reasoning models, recent studies~\cite{snell2024scaling,xu2025towards} suggest that longer reasoning chains often lead to better reasoning capabilities, thereby improving downstream task performance---this phenomenon is known as the \textbf{inference-time scaling law}. However, our experiments (\emph{cf.} Sec.~\ref{sec:performance_variation}) demonstrate that as the number of reasoning steps increases, our framework does not achieve the expected scaling law behavior in a perfect manner. This discrepancy raises several intriguing research questions: 
  \begin{itemize}
    \item \textit{Does a scaling law exist for inference-time computation in recommendation systems?}
    \item \textit{If so, how can we design more effective reasoning-enhanced sequential recommenders to better realize such a scaling law?}
  \end{itemize}
  Further exploration in this direction could unlock new insights into the model's reasoning capabilities and ultimately push the boundaries of reasoning-enhanced recommendation research.
  \item \textbf{\textit{Theoretical Analysis.}} Intuitively, increasing inference-time computational depth allows sequential recommenders to capture higher-order sequential feature crossing, leading to more accurate user preference predictions. To solidify this intuition, future work could focus on theoretical analysis of how multi-step reasoning contributes to improved recommendation performance.
  Establishing a strong theoretical foundation for reasoning-enhanced sequential recommendation could pave the way for more principled model design and optimization strategies.
  \item \textbf{\textit{Efficient Inference Mechanism.}} While our efficiency experiments~\ref{sec:reasoning_time} confirm that ReaRec introduces only marginal latency overhead, future advancements in sequential recommendation inference-time scaling laws may still raise efficiency concerns with the autoregressive generation paradigm. To address this, we propose several potential optimization strategies for future exploration, including incorporating linear attention mechanisms~\cite{wang2020linformer}, model quantization~\cite{yao2022zeroquant}, and long-to-short reasoning distillation~\cite{team2025kimi} techniques to further achieve lighter and faster inference efficiency for industrial-scale deployment.
\end{itemize}



\bibliographystyle{ACM-Reference-Format}
\bibliography{sample-sigconf}


\end{document}